\documentclass[aps,pre,twocolumn,superscriptaddress,showpacs]{revtex4-1}
\usepackage{graphicx}
\usepackage{array}
\usepackage{amsfonts}
\usepackage{amssymb,amsmath,multirow,rotate}
\usepackage{mathrsfs}
\usepackage{booktabs}
\usepackage{threeparttable}
\usepackage{multirow}
\usepackage{epsfig}
\usepackage{threeparttable}
\usepackage{chngpage}
\usepackage{latexsym}
\usepackage{dcolumn}
\usepackage{graphics,graphicx,bm,fleqn,epic,eepic,float}
\usepackage{verbatim}
\usepackage{subfigure}
\usepackage{color}
\usepackage{xr}

\definecolor{red}{rgb}{1,0,0}

\definecolor{yellow}{rgb}{1,0.5,0}
\definecolor{blue}{rgb}{0,0,1}

\begin{document}

\title{Scaling law of transient lifetime of chimera states under dimension-augmenting perturbations}
\date{\today}

\author{Ling-Wei Kong}
\affiliation{School of Electrical, Computer and Energy Engineering, Arizona State University, Tempe, Arizona 85287, USA}

\author{Ying-Cheng Lai} \email{Ying-Cheng.Lai@asu.edu}
\affiliation{School of Electrical, Computer and Energy Engineering, Arizona State University, Tempe, Arizona 85287, USA}
\affiliation{Department of Physics, Arizona State University, Tempe, Arizona 85287, USA}

\begin{abstract}

Chimera states arising in the classic Kuramoto system of two-dimensional phase coupled oscillators are transient but they are ``long'' transients in the sense that the average transient lifetime grows exponentially with the system size. For reasonably large systems, e.g., those consisting of a few hundreds oscillators, it is infeasible to numerically calculate or experimentally measure the average lifetime, so the chimera states are practically permanent. We find that small perturbations in the third dimension, which make system ``slightly'' three-dimensional, will reduce dramatically the transient lifetime. In particular, under such a perturbation, the practically infinite average transient lifetime will become extremely short, because it scales with the magnitude of the perturbation only logarithmically. Physically, this means that a reduction in the perturbation strength over many orders of magnitude, insofar as it is not zero, would result in only an incremental increase in the lifetime. The uncovered type of fragility of chimera states raises concerns about their observability in physical systems.

\end{abstract}

\maketitle

\section{Introduction}\label{sec:Intro}

A research frontier in complex and nonlinear dynamical systems is
chimera states~\cite{UGOA:1989,KB:2002,AS:2004,SK:2004,
AS:2006,AMSW:2008,SSA:2008,Laing:2009a,SCL:2009,
Martens:2010,MLS:2010, OWM:2010,WO:2011,WOYM:2011,OMHS:2011,OWYMS:2012,
ZLZY:2012,LRK:2012,TNS:2012,HMRHOS:2012,OOHS:2013,UR:2013,ZZY:2013,YHLZ:2013,
NTS:2013,MTFH:2013,LBM:2013,PA:2013,GSD:2013,SOW:2014,SSKG:2014,ZZY:2014,
O:2013,OMT:2008,XKK:2014,YHGL:2015,MSOM:2015,PA:2015,PAb:2015,MB:2015,
OZHSS:2015,BZSL:2015,NTS:2016,VA:2016,HBMR:2016,GF:2016,SZMS:2016,KHSKK:2016,
UOZS:2016,HBMR:2016,ARM:2017,BMGP:2017,RBPG:2017,SSAZ:2017,MOSH:2018,BK:2018,
OOZS:2018,XWHL:2018,YHRGL:2019}, a phenomenon of spontaneous symmetry
breaking in spatially extended systems in which coherent and incoherent groups 
of oscillators coexist simultaneously. The phenomenon was first observed three 
decades ago in a numerical study of the system of coupled nonlinear Duffing
oscillators~\cite{UGOA:1989}, which was later rediscovered~\cite{KB:2002}, 
analyzed and coined with the term ``chimera''~\cite{AS:2004,SK:2004}. Chimera 
states have been studied in diverse systems such as regular networks of 
phase-coupled oscillators with a ring topology~\cite{KB:2002,AS:2004,AS:2006}, 
networks hosting a few populations~\cite{AMSW:2008,Martens:2010},
two-dimensional (2D)~\cite{SK:2004,MLS:2010} and three-dimensional (3D)
lattices~\cite{MSOM:2015}, torus~\cite{PA:2013,OWYMS:2012}, and systems with 
a spherical topology~\cite{PA:2015}. Phenomena such as traveling-wave type
of chimera~\cite{XKK:2014} and amplitude chimera~\cite{BZSL:2015,HBMR:2016}
have also been uncovered and studied. 

In the Kuramoto model with 2D rotational dynamics, a previous 
study~\cite{WO:2011} demonstrated that the chimera states are typically 
transient. These states were deemed ``long transient'' because their average 
lifetime increases exponentially with the system size. For systems of size 
greater than, say, 60, it is already infeasible to numerically calculate the 
average transient time. For larger systems consisting of, e.g., a few hundreds 
oscillators, the average transient lifetime is practically infinite. The 
questions to be addressed in this paper are whether the practically infinitely 
long transient will become short so that the chimera states are actually 
transient from the standpoint of numerical computations or physical 
experiments when external perturbations are applied to the system, and how.
Previous work focused mostly on
perturbations to the structure of the underlying lattice or
networks~\cite{YHLZ:2013,OPHSH:2015,MOSH:2018}, revealing that chimera states
are robust against structural perturbations. For example, when
some links are removed from an originally globally coupled (all-to-all)
network, coherent and incoherent regions still simultaneously arise in
the state space, giving rise to a generalized type of chimera
states~\cite{YHLZ:2013}. Quite recently, it was found that a chimera
state can respond to perturbations to achieve robustness through the
mechanism of self-organization and adaptation~\cite{YHRGL:2019}.

In this paper, we report an unexpected type of fragility of chimera states
in the presence of perturbations to the phase-space dimension of the 
oscillators in the network. We start from the paradigmatic Kuramoto model of 
globally coupled 2D phase oscillators~\cite{KB:2002, AS:2004}. In this model, 
each oscillator is a 2D rotor characterized by a single dynamical variable, 
the angle of planar rotation. We invoke arbitrarily small perturbations that 
make the oscillator ``slightly'' 3D. Specifically, consider 3D rotation 
represented by the movement of a point on the surface of a unit sphere $S^2$, 
where 2D rotation of the unperturbed phase oscillator is confined to 
movements on the equator. 
We find that any infinitesimal deviation from the equator in the oscillator
dynamics makes the long-transient chimera state extremely short. In particular,
let $\delta$ be the strength of this kind of ``dimension-augmenting'' 
perturbation. We find that, regardless of how infinitesimal $\delta$ is,
insofar as its value is not zero, the average transient lifetime
$\langle\tau\rangle$ of the chimera states becomes extremely short as it
depends only logarithmically on $\delta$:  
$\langle\tau\rangle \sim -\ln{\delta}$, even for large systems for which
the average transient lifetime in 2D is practically infinite. The physical 
significance is that, when the strength of the perturbation is reduced 
by many orders of magnitude, the average transient lifetime would incur only 
an incremental increase.
Considering that in many existing studies of chimera states, whether it be 
physical, chemical or biological, a description based on Kuramoto type of 2D 
rotors is only approximate and perturbations that alter the 2D picture are 
inevitable, our finding raises concerns about the physical observability of 
chimera states.

It should be noted that, in this paper, an $N$-dimensional chimera state
for $N \ge 3$ is defined as one that emerges in the full $N$-dimensional
phase space as the result of dimension-augmenting perturbations. Because
the focus of our study is on the transient nature of such high-dimensional
chimera states, we set the initial condition to be a chimera state in two
dimensions as in the classical Kuramoto model and examine how long the state
can survive under such perturbations. This is done for two cases: the
perturbations are such that the local phase space of each oscillator becomes
three- or four-dimensional, respectively.
We also note that a previous work~\cite{LCSZ:2016} revealed a logarithmic 
dependence of the average lifetime of transient chimera states on the 
intensity of Gaussian white noise, indicating a dramatic reduction of the 
chimera lifetime under noise. Our finding of a similar scaling law but with 
respect to deterministic, dimension-augmenting perturbations is further 
indication of the fragility of chimera states.

\section{Model} \label{sec:Model}

\subsection{General consideration}

We begin with the following $D$-dimensional Kuramoto 
model~\cite{Olfati:2006,Lohe:2009,CGO:2019,CGO:2019b}:
\begin{align} \label{eq:HighDKuramoto}
\frac{d\boldsymbol{\sigma}_i}{dt}= \frac{K}{N}\sum_{j=1}^N\left[\boldsymbol{\sigma}_j-\left(\boldsymbol{\sigma}_j\cdot\boldsymbol{\sigma}_i\right)\boldsymbol{\sigma}_i\right]+\boldsymbol{{\rm W}}\cdot \boldsymbol{\sigma}_i,
\end{align}
where the $D$-dimensional unit vector $\boldsymbol{\sigma}_i$ represents the
state of the $i$th oscillator, $\boldsymbol{W}$ is a real $D\times D$
antisymmetric matrix characterizing the natural rotation of the oscillator,
and $N$ is the system size (the number of coupled oscillators).
The state of each node has $\left(D-1\right)$ degrees of freedom. For
$D=2$, the system reduces to the classic Kuramoto model with the variable 
substitution $\boldsymbol{\sigma}_i=\left(\cos{\theta_i},\sin{\theta_i}\right)$.
In Eq.~(\ref{eq:HighDKuramoto}), global (all-to-all) coupling is assumed, 
which is idealized. In a realistic situation, coupling is neither global nor 
highly local. We thus consider the following generalized model:
\begin{eqnarray} 
\nonumber
\frac{d\boldsymbol{\sigma}_i}{dt} & = & \frac{1}{N}\sum_{j=1}^NG\left( i-j \right)\left[\boldsymbol{{\rm T}}\cdot\boldsymbol{\sigma}_j-\left((\boldsymbol{{\rm T}\cdot\sigma}_j)\cdot\boldsymbol{\sigma}_i\right)\boldsymbol{\sigma}_i\right] \\ \label{eq:OurModel}
& + & \boldsymbol{{\rm W}}\cdot\boldsymbol{\sigma}_i,
\end{eqnarray}
where $G\left( i-j \right)$ is a coupling function of a finite range, e.g.,
$G\left( i-j \right)=1+A \cos{\left[2\pi\left(i-j\right)/N\right]}$, and 
$\boldsymbol{{\rm T}}$ is a $D\times D$ isometric matrix taking into account
phase lag. The oscillators can be visualized to be located 
on a ring and the coupling strength between a pair of nodes decreases with 
their distance according to $G(i-j)$. The state vector $\boldsymbol{\sigma}_i$ 
is now a high-dimensional unit vector. Let the starting point of 
$\boldsymbol{\sigma}_i$ be the origin so its ending point is on the surface 
of the high-dimensional unit sphere. In 3D the system can be conceived as a 
``pearl necklace,'' as shown in Fig.~\ref{fig:TCS}(a), where 
$\boldsymbol{\sigma}_i$ of each oscillator moves on the surface of its pearl 
and all the pearls are located on the ring.

In 2D, the isometric matrix $\boldsymbol{{\rm T}}$ reduces to the standard 
rotation matrix of angle $\alpha$ as
\begin{align}
\boldsymbol{{\rm T}}=
\begin{bmatrix}
\cos\alpha & \sin\alpha \\
-\sin\alpha & \cos\alpha \\
\end{bmatrix}.
\end{align}
While acting on a vector, \boldsymbol{{\rm T}} alters its direction 
while preserving its length, thus serves as a phase lag. Given an isometry in 
$d$ dimensions, there exists a reference framework in which 
$\boldsymbol{{\rm T}}$ can be written in the form
\begin{align}
\boldsymbol{{\rm T}}=
\begin{bmatrix}
\boldsymbol{{\rm PR_1}} \\
&\boldsymbol{{\rm PR_2}} \\
& &{...} \\
& & &\boldsymbol{{\rm PR_k}} \\
& & & &\boldsymbol{{\rm RF_m}} \\
& & & & &\boldsymbol{{\rm I_{d-2k-m}}}
\end{bmatrix},
\end{align}
where $\boldsymbol{{\rm PR_i}}$ ($i=1,2...k$) is the proper rotation matrix
in 2D:
\begin{align}
\boldsymbol{{\rm PR_i}}=
\begin{bmatrix}
\cos\alpha_i & \sin\alpha_i \\
-\sin\alpha_i & \cos\alpha_i \\
\end{bmatrix},
\end{align}
$\boldsymbol{{\rm I_{d-2k-m}}}$ is the$(d-2k-m)\times(d-2k-m)$ identity matrix,
and $\boldsymbol{{\rm RF_m}}$ is the $m\times m$ reflection matrix with all 
the diagonal elements $-1$ and all the off-diagonal elements zero:
$\boldsymbol{{\rm RF_m}}=-\boldsymbol{{\rm I_m}}$. We can use the three 
integers above: $d$, $k$ and $m$, to classify all different types of 
isometry subject to the constraint $d\geq2k+m$. We impose another constraint: 
$k\geq 1$, to ensure a finite phase lag.

For $d=2$, the only choice is $k=1$ and $m=0$, which is simply the 2D proper
rotation matrix. For $d=3$, it is necessary to choose $k=1$, and $m$ can be
either $1$ or $-1$. We study both cases. For $d=4$, we have four different
choices: (1) $k=2$ and $m=0$, (2) $k=1$ and $m=0$, (3) $k=1$ and $m=1$, and
(4) $k=1$ and $m=2$.

With the following space and time dependent order parameter:
$\boldsymbol{\rho}_i=N^{-1}\sum_{j=1}^NG\left(i-j\right)\boldsymbol{\sigma}_j$,
we rewrite our generalized $D$-dimensional Kuramoto model as
\begin{align} \label{eq:OurModelOrderParameter}
d\boldsymbol{\sigma}_i/dt=\boldsymbol{{\rm T}\rho}_i-\left(\boldsymbol{{\rm T}\rho}_i\cdot\boldsymbol{\sigma}_i\right)\boldsymbol{\sigma}_i+\boldsymbol{{\rm W}}\boldsymbol{\sigma}_i.
\end{align}

\subsection{Articulation of dimension-augmenting perturbations}

To be concrete, we focus on perturbations that make the system 3D. To 
investigate the lifetime of chimera states, we articulate a scheme such that 
the 2D chimera states are a natural solution of the system. We then perturb 
this solution into 3D and determine whether or not it is still a long 
transient. (It should be emphasized that, without any perturbation, given the 
symmetry of the system about the $(x,y)$-plane, the system would remain 2D, 
and all the results would be the same as for the 2D system.)

More specifically, for vector
rotation on a sphere, there are two independent dynamical variables: the
longitudinal and latitudinal angles. For a network of size $N$, the phase
space dimension is thus $2N$. For oscillator $i$, let $0\leq\theta_i<2\pi$
and $-\pi/2\leq\gamma_i\leq\pi/2$ be the longitudinal and latitudinal angles,
respectively. To make the $N$-dimensional subspace defined by $\gamma_i = 0$
($i=1,\ldots,N$) an invariant subspace of the system, we choose a frame in
which the initial 2D plane is the plane containing the equator and set the
$z$-axis of the frame to be the axis of the proper rotation component of
$\boldsymbol{{\rm T}}$. In this frame, the isometric matrix 
$\boldsymbol{{\rm T}}$ is:
\begin{align}
\boldsymbol{{\rm T}} =
\begin{bmatrix}
\cos{\alpha} & \sin{\alpha} &0 \\
-\sin{\alpha} & \cos{\alpha} &0 \\
0 &0 & T_{33}
\end{bmatrix},
\end{align}
According to our definition of $\boldsymbol{{\rm T}}$ in Sec.~\ref{sec:Model}, 
$T_{33}$ can be either $1$ or $-1$, corresponding to whether a reflection 
symmetry is excluded or included, respectively. 
For $T_{33}=1$, no chimera state can arise because the system dynamics are 
such that all oscillators quickly synchronize to the fixed points of the
transformation $\boldsymbol{{\rm T}}$: $(0,0,\pm 1)$ (in the Cartesian 
coordinates). We thus focus on the case $T_{33}=-1$ here and treat the case 
$T_{33}=1$ in Sec.~\ref{sec:T33And4D}.

We make the natural rotation $\boldsymbol{{\rm W}}$ about the $z$-axis, so the 
system degenerates to 2D in the absence of any perturbation. In this case, 
$\boldsymbol{{\rm W}}$ plays no role in the dynamics, since it can be removed
by setting the reference frame to one rotating about the z-axis at the same 
frequency. These considerations lead us to rewrite the general system equation 
Eq.(\ref{eq:OurModelOrderParameter}) in the spherical coordinate as
\begin{eqnarray} \label{eq:SphericalTheta}
\frac{d\theta_i}{dt} & = & -R_i\frac{\cos\Gamma_i}{\cos\gamma_i}\sin\left(\theta_i-\Theta_i+\alpha\right) \\ \nonumber
\label{eq:SphericalGamma}
\frac{d\gamma_i}{dt} & = & -R_i\times [\cos\left(\theta_i-\Theta_i+\alpha\right)\sin\gamma_i \cos\Gamma_i \\
& - & T_{33}\cos\gamma_i\sin\Gamma_i]
\end{eqnarray}
where $\Theta_i$, $\Gamma_i$ and $R_i$ are the longitudinal angle, latitudinal
angle and the length of the order parameter $\boldsymbol{\rho}_i$ at the 
location of the $i$th oscillator, respectively. Note that, in 2D, the equation 
for the single dynamical variable $\theta_i$ is
\begin{displaymath}
d\theta_i/dt=-R_i\sin\left(\theta_i-\Theta_i+\alpha\right). 
\end{displaymath}	
Comparing this with Eq.~(\ref{eq:SphericalTheta}), we see that the extra 
factor in 3D is 
\begin{equation} \label{eq:dimensionality_measure}
\chi_i \equiv \cos\Gamma_i/\cos\gamma_i, 
\end{equation}	
which we name as the dimensionality measure. For $\gamma_i = 0$ 
($i=1,\ldots,N$), we have $\Gamma_i = 0$ and $\chi_i = 1$, so Eq.~(\ref{eq:SphericalTheta}) 
reduces to the 2D form, meaning that the 2D chimera states are an invariant 
solution of the 3D system.

\section{Scaling results} \label{sec:results}

\subsection{Logarithmic scaling of average transient lifetime of chimera 
states with dimension-augmenting perturbation in 3D}

\begin{figure}[htp!]
\centering
\includegraphics[width=\linewidth]{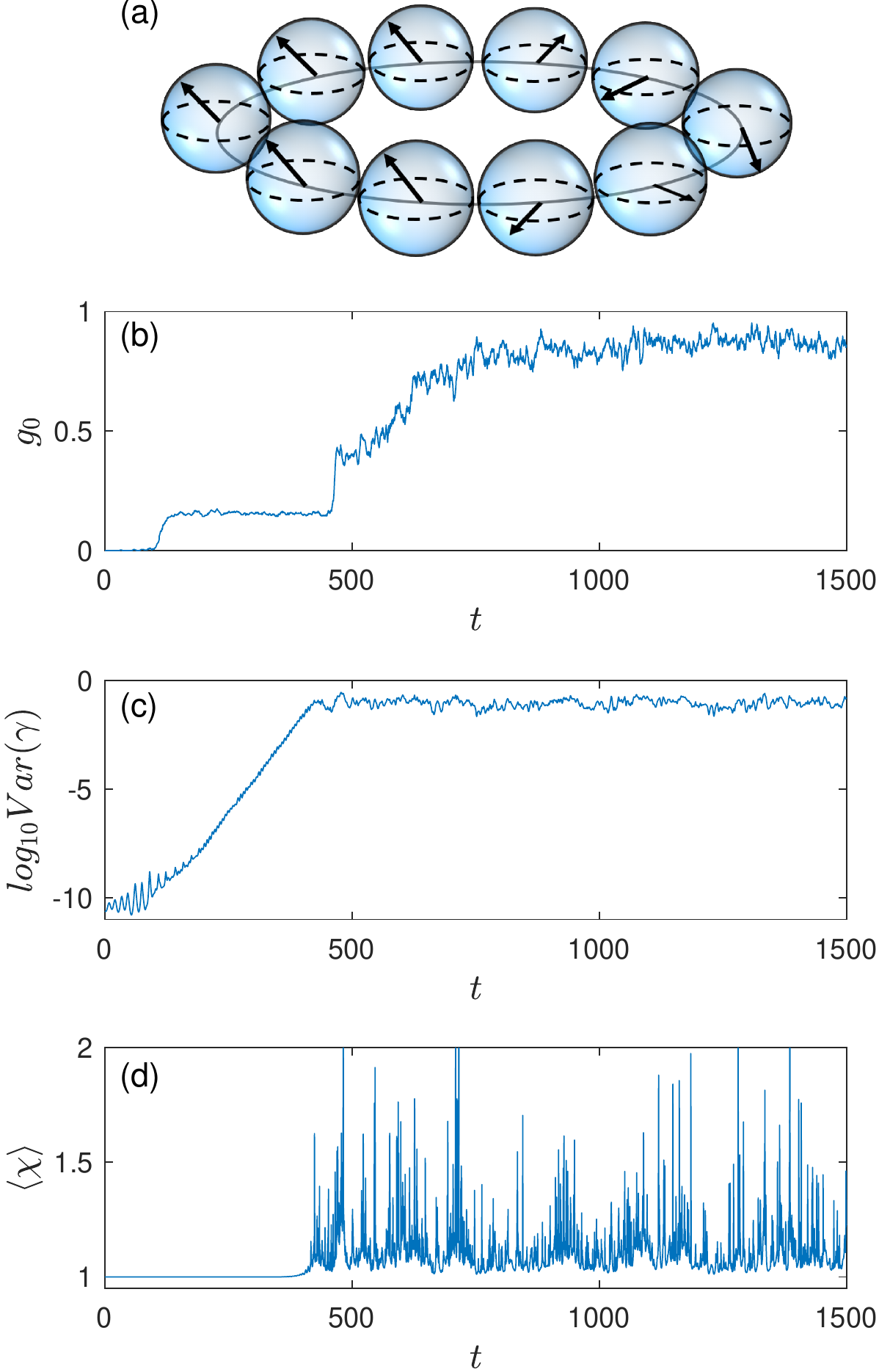}
\caption{ {\em System illustration and transient chimera states in 3D}.
(a) The system can be conceived as a ``pearl necklace'' with each oscillator 
oscillating on the surface of its $3D$ ``pearl'' and all the pearls being 
located on a ring. (b) Time evolution of $g_0\left(t\right)$. The time 
interval in which the relative size $g_0\left(t\right)$ of the coherent 
region in space is approximately constant indicates the existence of a 
chimera state. Variations in $g_0\left(t\right)$ arise for $t > 459$, 
signifying the disappearance of the 2D-like chimera state. (c) Evolution 
of $\mbox{Var}(\gamma)$, the variance of $\gamma$, which increases 
approximately exponentially during the transient 2D-like chimera state and 
reaches a plateau a short time before the collapse of the 2D-like structure. 
(d) Time evolution of the space averaged dimensionality measure 
$\langle \chi_i\rangle$ in Eq.~(\ref{eq:SphericalTheta}). For $t \alt 400$, 
the value of $\langle \chi_i\rangle$ remains at one, indicating that the 
network dynamics are essentially of the 2D Kuramoto type. Parameter values 
are $N=400$, $T_{33}=-1$, $\delta=0.0001$, $A=0.995$, and $\alpha=\pi/2-0.05$.
We use this pair of values of $A$ and $\alpha$ because the basin of the 2D 
chimera state is relatively large, facilitating numerical simulations.}
\label{fig:TCS}
\end{figure}

A previous study~\cite{WO:2011} of the 2D version of
Eq.~(\ref{eq:OurModelOrderParameter}) indicated the existence of transient
chimera states with a long lifetime. The transient time increases 
exponentially with the system size $N$, making numerical simulations 
infeasible to observe the collapse of the chimera state for, e.g., $N>60$. 
Our question is whether the chimera states can sustain such a long lifetime 
when the 2D system is perturbed into a higher-dimensional one.

To detect possible chimera states in 3D, we calculate the discrete Laplacian
$D_i$ at node $i$ as a measure of the spatial coherence and derive the relative 
size $g_0$ of the coherence region in the space~\cite{KHSKK:2016}. We then 
calculate the time evolution of $g_0$ and the instantaneous distributions of
$D_i$. A finite time interval in which $g_0$ is approximately constant while
the distribution of $D_i$ has two peaks signifies the existence of a transient
chimera state (Appendix~\ref{sec:SMDistChimeraState}). Figure~\ref{fig:TCS}(b) 
shows a typical behavior of the time evolution of $g_0$, where its value 
increases from zero initially and reaches a plateau at $t \approx 120$. For 
$120 \alt t \alt 460$, $g_0$ is constant. For $t \agt 460$, the value of $g_0$ 
increases continuously, indicating a deterioration of the chimera state and 
system's approaching a global, loosely synchronous state
(Sec.~\ref{sec:SMCollapse}). 

The transient nature of the observed chimera state can be understood, as 
follows. Start from a random set of 2D initial conditions for the 
oscillators, i.e., $\gamma_i(0) = 0$. Without any perturbation, the system 
dynamics remain 2D with $\gamma_i(t) = 0$ for all $t$. In this case, chimera 
states of long duration can arise insofar as the system size is not too 
small~\cite{WO:2011}. However, with a small perturbation to the latitudinal 
angle of a single oscillator in the initial condition, e.g., 
$\delta = 10^{-3}$ in Fig.~\ref{fig:TCS}(b), the system will remain to be 
approximately 2D for only a finite amount of time, which can be seen from the 
behaviors of the space averaged values of the variance of $\gamma_i$ and 
of the dimensionality measure $\langle\chi_i\rangle(t)$, as shown in 
Figs.~\ref{fig:TCS}(c) and \ref{fig:TCS}(d), respectively. These results 
demonstrate that, for $t \alt 450$, the dynamics of the oscillators are
effectively 2D but they become 3D afterwards. 

\begin{figure}[htp!]
\centering
\includegraphics[width=\linewidth]{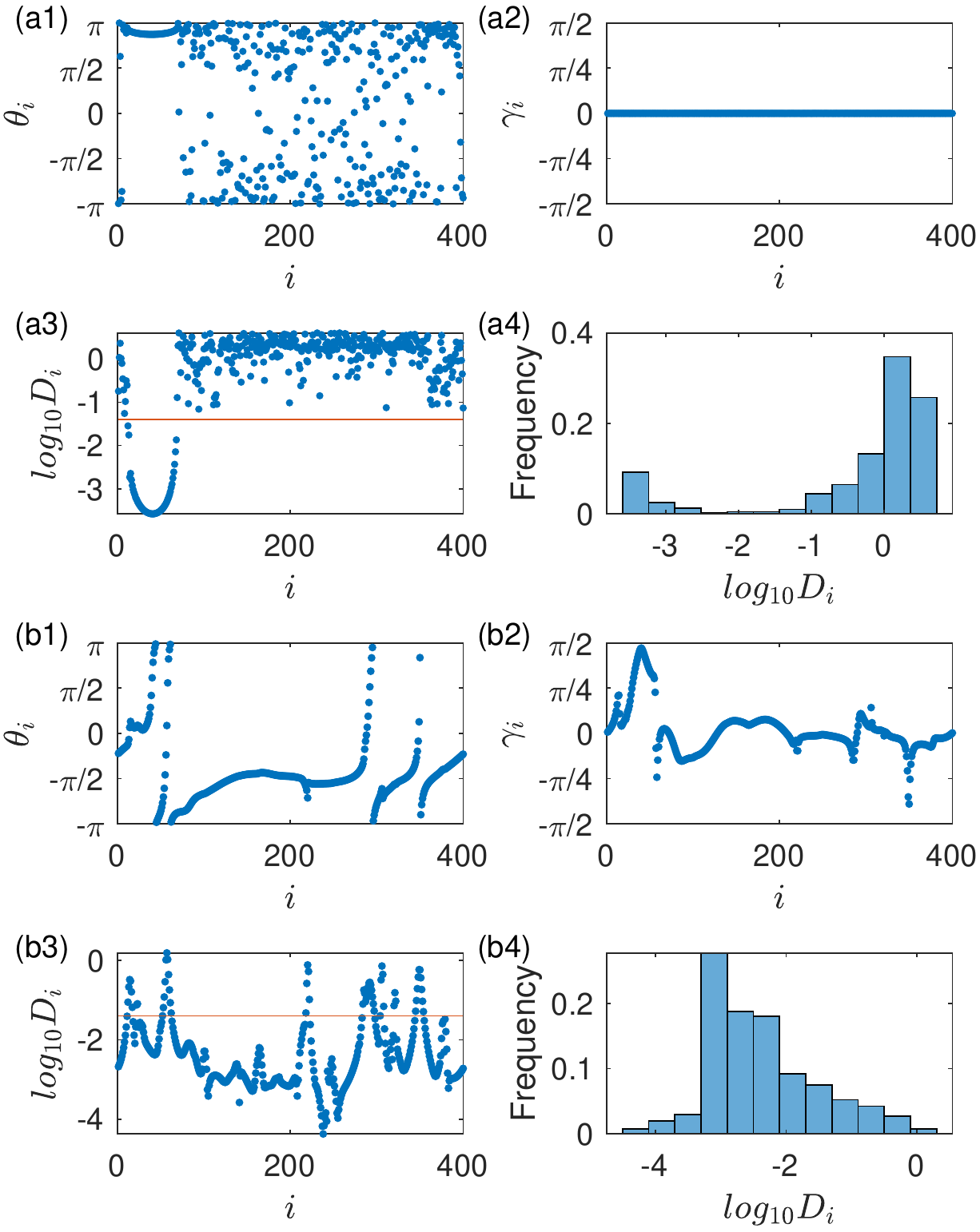}
\caption{ {\em Visualization of system states from Fig.~\ref{fig:TCS}}.
(a1-a4) Snapshots at $t=200$ when the system is still in a 2D-like chimera
state. (b1-b4) Snapshots at $t=1500$ when the system has left the 2D
state. (a1,b1) Longitudinal angles $\theta_i$ of
all the oscillators at the two time instants. (a2,b2) The latitudinal angles 
$\gamma_i$. (a3,b3) The values $\log_{10}D_i$ of all the oscillators, where
$D_i$ is the spatial Laplacian of $\boldsymbol{\sigma}$ and characterizes
the instant local degree of distortion~\cite{KHSKK:2016}. If $D_i$ is below a 
threshold (e.g., $0.04$), as shown by the red horizontal line, oscillator $i$ 
is within the coherence region (see Appendix~\ref{sec:SMDistChimeraState} 
for more details). (a4,b4) Histograms of $\log_{10}{D_i}$ at
the two time instants. For $t=200$, in panel (a4) there are two peaks: one 
at a high and another at a low value of $\log_{10}{D_i}$, indicating 
coexistence of incoherent and coherent regions. However, for $t=1500$, 
as shown in panel (b4), only one peak stands, signifying the disappearance 
of the chimera state.}
\label{fig:snapshot}
\end{figure}

Snapshots of the chimera state are presented in 
Figs.~\ref{fig:snapshot}(a1-a4), while those after the state has disappeared 
are shown in Figs.~\ref{fig:snapshot}(b1-b4). As shown in 
Fig.~\ref{fig:snapshot}(a4), two peaks arise in the distribution of $D_i$, 
indicating a chimera state in the time interval $120 \alt t \alt 460$. 
For $t \agt 460$, the coherence distribution has only one peak, signifying
a globally synchronous state, as shown in Fig.~\ref{fig:snapshot}(b4). The 
remarkable phenomenon is that the chimera state occurs essentially during the 
time interval where the system is effectively 2D. As the dynamics become 3D, 
the chimera state deteriorates and disappears quickly. The behaviors
illustrated in Figs.~\ref{fig:TCS}(b-d) hold regardless of the specific
initial conditions. The message is that chimera states cannot survive against
perturbations that alter the 2D nature of the oscillator dynamics.

\begin{figure}[htp!]
\centering
\includegraphics[width=\linewidth]{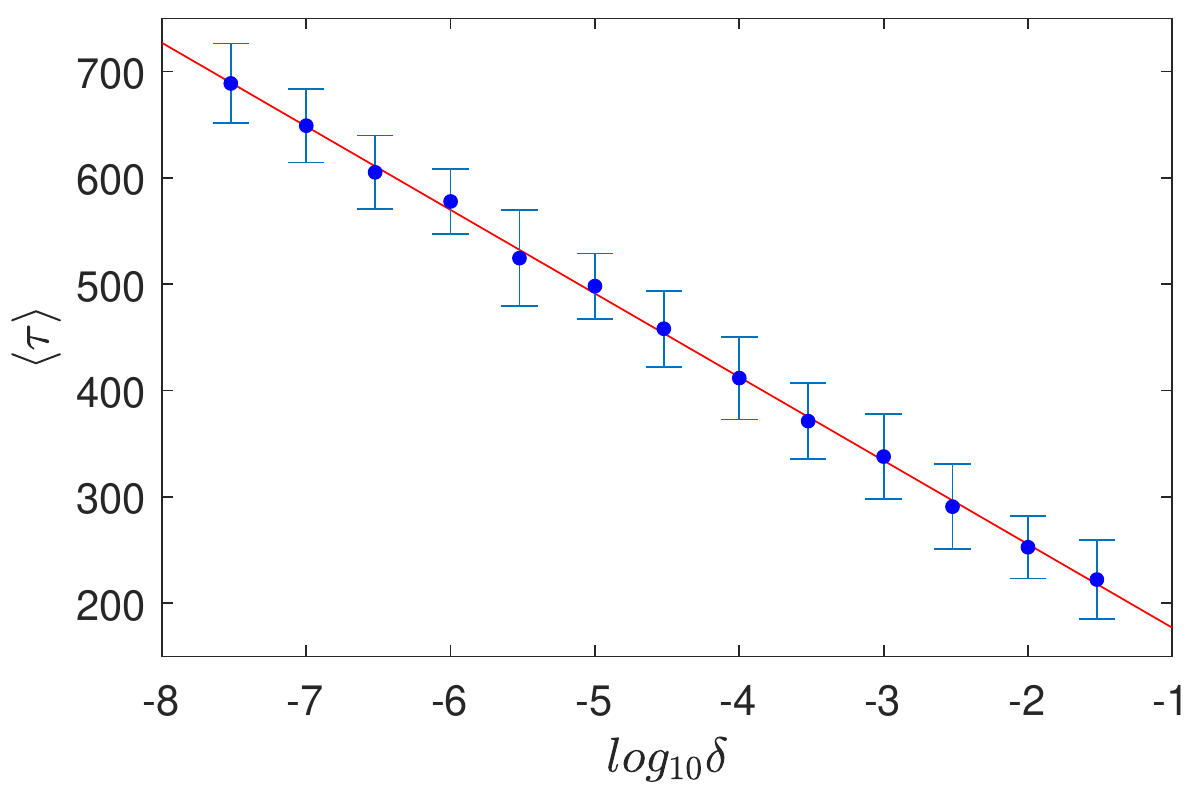}
\caption{ {\em Scaling of the average chimera lifetime with perturbation}.
A fit of the data points gives the scaling between $\langle\tau\rangle$
and the magnitude of the dimension altering perturbation $\delta$ as
$\langle\tau\rangle \sim - \log_{10}{\delta}$. Error bars represent standard
deviations of the distributions. Parameter values are $N=256$, $A=0.995$,
and $\alpha=\pi/2-0.05$.}
\label{fig:tvsgammad}
\end{figure}

For any transient phenomenon in nonlinear dynamics, a fundamental issue is the
scaling law of the average transient lifetime with some system parameter
variation, noise amplitude, or perturbation~\cite{GOY:1983,LT:book}. A
transient behavior will be physically equivalent to some attracting behavior
if the transient lifetime diverges exponentially, as speculated in the case
of turbulence~\cite{CK:1988} or superpersistent chaotic
transients~\cite{GOY:1985,DL:2003,DL:2004,DLHG:2016,LDHSG:2017}. Such scaling
was also found for chimera states in networks of Boolean phase 
oscillators~\cite{RRHSG:2014}. Under a dimension-augmenting perturbation, how 
does the average chimera time scale with the perturbation strength? A 
representative result is shown in Fig.~\ref{fig:tvsgammad}, where the average 
chimera time $\langle\tau\rangle$ (on a linear scale) is plotted against the 
perturbation magnitude $\delta$ (on a logarithmic scale). We have the scaling 
law: $\langle\tau\rangle \sim -\ln{\delta}$, the physical significance of 
which is that the transient lifetime is {\em extraordinarily short}, in 
contrast to many transient scaling laws in nonlinear dynamical 
systems~\cite{LT:book}. In fact, a reduction in the perturbation by many 
orders of magnitude results in only an incremental increase in the average 
chimera time. That is, an arbitrarily small perturbation that drives the 
oscillator dynamics away from 2D immediately destroys the long-transient 
chimera state.

Why does a chimera state collapse when the oscillator dynamics deviate away 
from 2D? The value of $R_i\chi_i$ is the key. In 3D, the dynamics of 
$\theta$ are governed by Eq.~(\ref{eq:SphericalTheta}), where the only 
difference with the 2D model is the extra factor $\chi_i$. While it
appears that, in 3D, the value of $R_i\chi_i$ may play a similar rule to that 
of $R_i$ in 2D, $R_i\chi_i$ is affected not only by the coherence of the 
oscillators about the $i$th oscillator, but also by the latitudinal angles 
of all the oscillators, especially $\gamma_i$. Typically, the value of 
$\Gamma_i$ is close to zero, but $\gamma_i$ can be away from zero. In 
2D, since $R_i$ is a measure of coherence, its value for an oscillator in
the incoherent region must be smaller than that in the coherent region. 
However, in 3D, incoherent oscillators can have larger $R_i\chi_i$ values 
than those of the coherent ones, because the former are less coherent and can 
diffuse away from the initial equator faster, resulting in larger values of 
$|\gamma_i|$. Similar to the role of large values of $R_i$ in 2D,  large 
values of $R_i\chi_i$ in 3D will make the oscillators more coherent. 
Simulations have revealed (Sec.~\ref{sec:SMCollapse}) that
$R_i\chi_i$ can be large for many oscillators in the incoherent region,
leading to the emergence of a new and wider coherent region within the 
incoherent region. After that, an increasing number of coherent regions 
form inside the remaining incoherent regions, making the whole system closer 
to being globally coherent.

The collapse of the 2D-like structure is then caused by the diffusion 
of oscillators in their $\gamma$ components. When the $\gamma_i$ values of 
some oscillators in the incoherent region are not close to zero, a new 
coherent region is formed. How far away from the initial equator the $\gamma_i$
values collectively are can be measured by the variance of $\gamma_i$, as 
shown in Fig.~\ref{fig:TCS}(c). We see that $\mbox{Var}(\gamma)$ 
tends to increase exponentially during the 2D-like chimera state, leading to 
the observed logarithmic dependence of the average lifetime on the 
perturbation strength. In particular, let $v_{max}$ be the threshold of 
$\mbox{Var}(\gamma)$ beyond which a new coherent region emerges
and let $\langle t_e\rangle$ be the average time that the threshold is 
reached. We have 
\begin{displaymath}
\delta \exp{(\kappa \langle t_e\rangle)} \sim v_{max}.
\end{displaymath}
Since $\langle\tau\rangle \sim \langle t_e\rangle$, we get $\langle\tau
\rangle \sim -\ln{\delta}$.

In the 2D system, the chimera states typically coexist with the complete  
synchronization state. This is the reason that we choose the values of $A$ and
$\alpha$ to be close to one and $\pi/2$, respectively. In this parameter 
regime, the basin of the chimera states is relatively large, facilitating 
numerical observation of a chimera state with random initial conditions. 
It is insightful to compare the results with those for the case where the 
initial conditions are chosen from the basin of the complete synchronization 
state. In this case, we have that, after a short transient, the system 
approaches an asymptotically global synchronization state. The length of
this transient is comparable to that of the transient before the emergence 
of the chimera states from initial condition in the chimera-state basin, 
which is about $t=120$ in Fig.~\ref{fig:TCS}(b). Upon a dimension-augmenting 
perturbation, the synchronous state of the system remains to be low-dimensional.
That is, the system will not become 3D. We thus see that the situation with
the chimera state is characteristically different: such a state becomes
3D but only for a short transient period of time before its collapse. This 
point can also be seen from Fig.~\ref{fig:TCS}(c) where, during the transient 
chimera phase, the system rapidly moves out of the equator, with an 
exponentially growing variance in the latitudinal angle $\gamma$.

\subsection{Dependence of transient lifetime of chimera states on system size} 
\label{sec:SMSystemSize}

\begin{figure}[htp!]
\centering
\includegraphics[width=\linewidth]{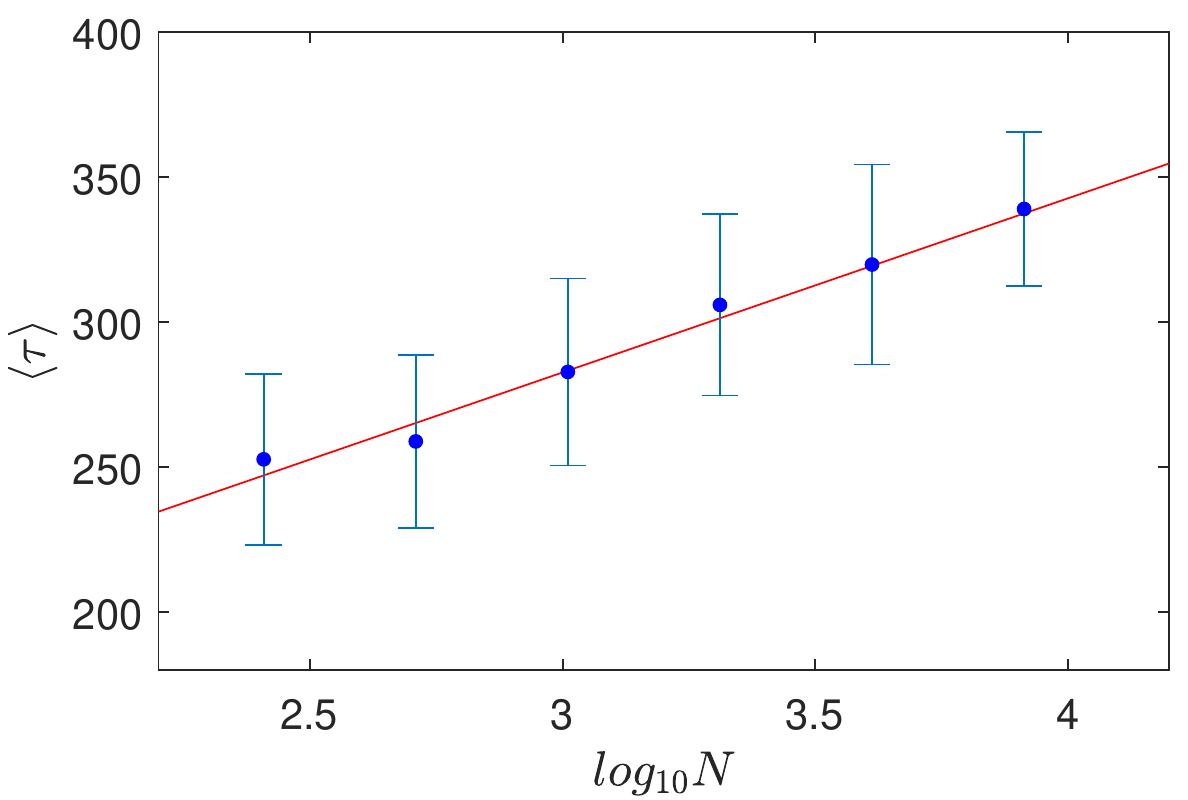}
\caption{ {\em Average transient lifetime of chimera states versus system size 
in 3D}. Shown are numerical results (dots) and linear fit (line). Error bars 
represent the standard deviations of the distributions. The system parameters 
are $T_{33}=-1$, $\theta_d=0.01$, $A=0.995$ and $\alpha=\pi/2-0.05$.}
\label{fig:tvsn}
\end{figure}

In 2D, the average lifetime of a chimera state grows exponentially with the 
system size, rendering infeasible numerical simulation~\cite{WO:2011} for 
large systems. However, we find that, in higher dimensions, the average 
chimera time does not follow such a rule, as the mechanism of the collapse 
of the chimera state is quite different from that in 2D. 
Figure~\ref{fig:tvsn} shows that the average chimera time scales with the 
system size only logarithmically. The heuristic reason is that an increase 
in the number of oscillators weakens, on average, the influence of the 
perturbation applied to a single node on other nodes. Since the average 
chimera time scales with the magnitude of the perturbation only 
logarithmically, so should be its dependence on the system size.

\begin{figure}[htp!]
\centering
\includegraphics[width=\linewidth]{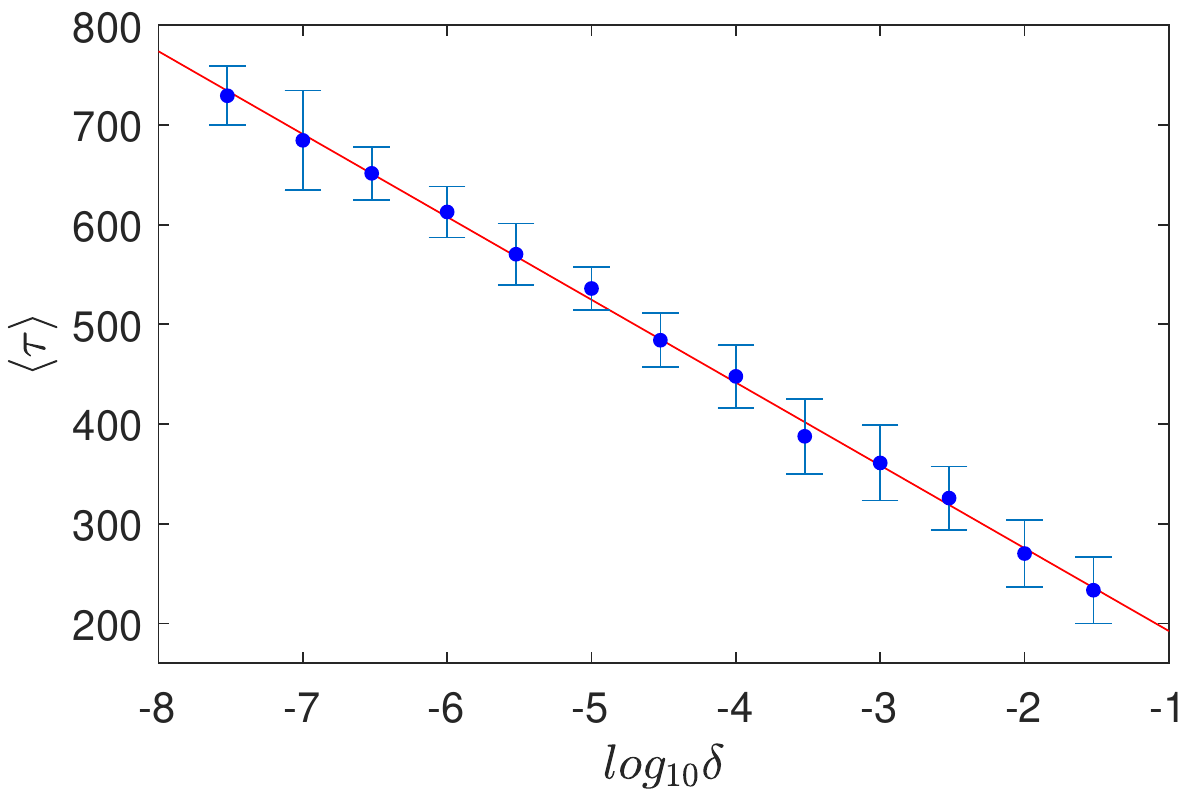}
\caption{ {\em Scaling of the average chimera lifetime with 
perturbation for a larger system size}. The system size is $N=1024$. 
The scaling law is essentially the same as that in Fig.~\ref{fig:tvsgammad} 
for $N=256$, with only about a $10\%$ increase in the average transient 
lifetime. Other parameter values are $T_{33}=-1$, $\theta_d=0.01$, $A=0.995$, 
and $\alpha=\pi/2-0.05$.}
\label{fig:5}
\end{figure}

To assess the generality and reliability of the uncovered scaling law of the
average chimera lifetime versus the dimension-augmenting perturbation, we
have calculated the scaling law for different values of the system size $N$. 
An example is presented in Fig.~\ref{fig:5}, where the scaling law is obtained 
for $N = 1024$. Comparing with the scaling law in Fig.~\ref{fig:tvsgammad} 
for $N = 256$, we see that a
four-fold increase in the system size does not change the logarithmic
scaling law. In fact, the only noticeable change is a slight increase in
the average lifetime (about $10\%$), due to the logarithmic nature of the
scaling law. This provides further support for our finding of the fragility
of the chimera state because a dramatic increase in the system size does
not significantly prolong the transient. This should be contrasted to the
case of chaotic transients in spatiotemporal dynamical systems, where the
average transient lifetime typically increases extremely rapidly with the
system size, often in a way that is faster than exponential 
growth~\cite{CK:1988}.

\subsection{Dependence of average transient lifetime of chimera 
states on coupling parameters} \label{sec:SMCouplingParameter}

\begin{figure*}[htp!]
\centering
\includegraphics[width=\linewidth]{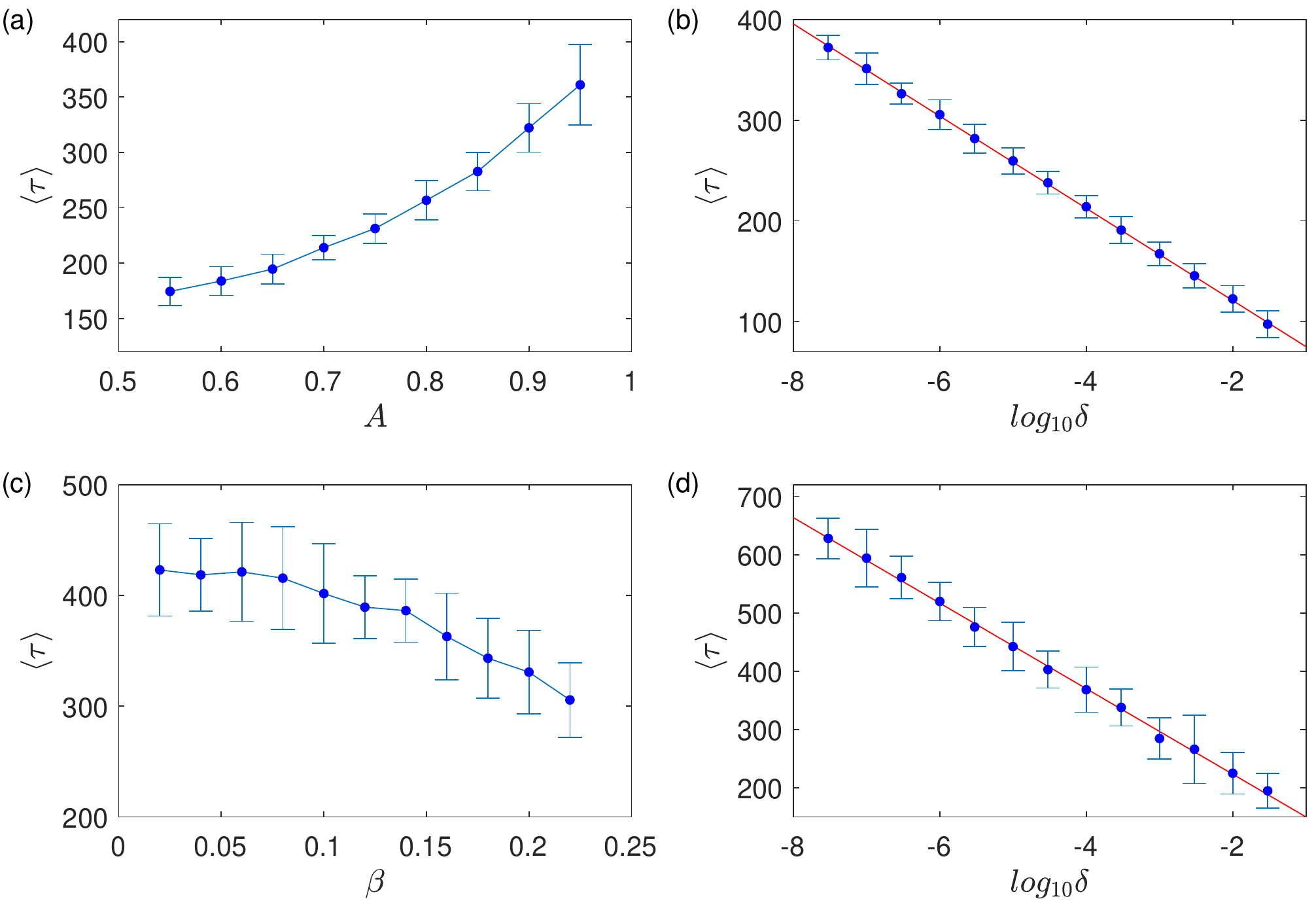}
\caption{ {\em Average transient lifetime of chimera states and 
scaling law for different values of the coupling parameters $A$ and $\beta$ 
in 3D}. (a) Average chimera lifetime versus $A$ for $T_{33}=-1$, 
$\theta_d=0.0001$, and $\beta=\pi/2-\alpha=0.05$. (b) A representative scaling
law of the average chimera lifetime with perturbation for $A=0.7$, where other 
parameter values are the same as those in (a). (c) Average chimera lifetime 
versus $\beta$ for $T_{33}=-1$, $\theta_d=0.0001$, and $A=0.995$. (d) A 
representative scaling law for $\beta=0.15$. Other parameters are the same 
as those in (c). In all panels, the error bars represent the standard 
deviations of the probability distributions.} 
\label{fig:6}
\end{figure*}

We study how the average transient lifetime of the chimera states depends
on the coupling parameters $A$ and $\alpha$. For convenience, we introduce
$\beta=\pi/2-\alpha$ to facilitate analysis of the situation where the value
of $\alpha$ is close to $\pi/2$ and that of $A$ close to one so as to obtain
a relatively large basin of the chimera states. In fact,
as the values of $\beta$ and $A$ deviate from zero and one, respectively,
the basin of the globally synchronized state will be enlarged, eventually
making the basin of the chimera states vanish~\cite{AS:2004}.

Figure~\ref{fig:6}(a) shows that the average transient lifetime 
$\langle\tau\rangle$ of the chimera states increases with $A$, with a 
representative scaling law for $A = 0.7$ shown in Fig.~\ref{fig:6}(b). 
Figure~\ref{fig:6}(c) shows the dependence of $\langle\tau\rangle$ on 
$\beta$, with the scaling law for $\beta = 0.15$ shown in Fig.~\ref{fig:6}(d). 
In general, when the values of $A$ and $\beta$ move
closer to the boundary beyond which chimera states no longer exist, the
average transient lifetime decreases, due to the system's transition into
3D as characterized by a faster growth of $Var(\gamma)$. In addition,
not only will the proportion of the initial states that go directly to
the complete synchronization state increase, but the final states after the
collapse of a transient chimera state will also be different (see the last 
paragraph of Sec.~\ref{sec:SMCollapse} for a further discussion of this 
phenomenon). Despite these behaviors, the logarithmic scaling law between 
the average transient time and perturbation remains invariant. These results, 
together with the results in Sec.~\ref{sec:SMSystemSize}, attest to the 
remarkable robustness of the scaling law.

\section{Mechanism of collapses of chimera states in high dimensions}
\label{sec:SMCollapse}

\begin{figure}[htp!]
\centering
\includegraphics[width=\linewidth]{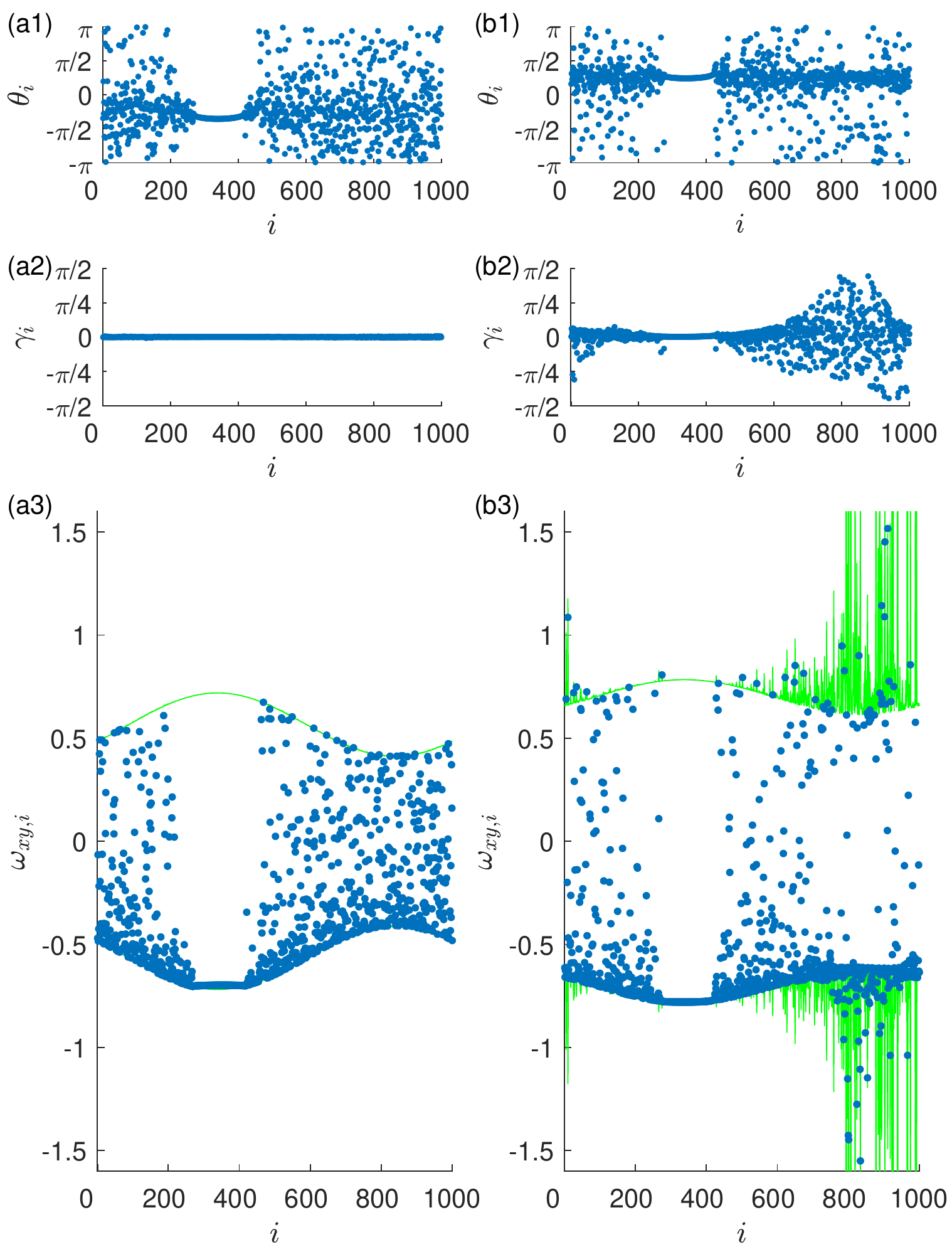}
\caption{ {\em Snapshots of the 3D system}. 
(a1, a2, a3) The first stage and (b1, b2, b3) the second stage as discussed 
in Sec.~\ref{sec:SMCollapse}. (a1, b1) Snapshots of the longitudinal angles 
$\theta_i$ of the oscillators. In both stages, the system is in a chimera state
in terms of component $\theta$. (a2, b2) Snapshots of the latitudinal angles 
$\gamma_i$ of the oscillators. In the first stage (a2), all $\gamma_i$ values 
are still close to zero, while in the second stage (b2) many oscillators close 
to the center of the incoherent region have $\gamma_i$ values away from zero. 
(a3, b3) Snapshots of the instantaneous angular velocity in the $(x,y)$-plane, 
$\omega_{xy,i}$ (blue dots), and $\pm R_i\chi_i$ (green traces) of all the 
oscillators. Equation~(\ref{eq:SphericalTheta}) gives that $\omega_{xy,i}$ is 
bounded by $\pm R_i\chi_i$, which agrees with the simulation results. In the 
first stage (a3), the coherent region has the largest and locked values of 
$\omega_{xy,i}$, similar to the regular 2D chimera state, and the oscillators 
in the incoherent region cannot reach such a value of $\omega_{xy,i}$ since 
they are bounded by the smaller values of $R_i\chi_i$. However, in the second 
stage (b3), $\pm R\chi$ exhibits irregular spikes for a large number of 
oscillators close to the center of the incoherent region, so these oscillators 
can have similar value of $\omega_{xy,i}$ to that of locked $\omega_{xy}$ 
in the incoherent region, or even larger. This indicates that the local 
dynamics of these oscillators have already deviated significantly from the 
2D-like structure.}
\label{fig:stage12}
\end{figure}

The general picture of the collapse of the chimera state can be described in 
terms of five successive stages, as follows. In the first stage, a 2D-like 
chimera state is formed with the coexistence of a coherent and an incoherent 
region. In this stage, the components of the state vectors 
$\boldsymbol{\sigma}$ of all the oscillators in the third dimension, 
$\boldsymbol{\sigma}_z$, are small. In the spherical coordinates, all 
oscillators have their $\gamma$ values close to zero, so the corresponding 
$\chi$ values are all close to one. The incoherent region has a lower $R$ 
value due to its lack of coherence, so the corresponding value of $R\chi$ is 
also small, making the oscillators there less affected by the collective 
behavior. In our approach, the initial state is that all the oscillators have 
$\gamma_i=0$, except for one oscillator with a small perturbation 
$\gamma_i=\delta$. During the system evolution, the values of $\gamma_i$ in 
the incoherent region gradually diffuse due to interactions, while $\gamma$ 
in the coherent region remains at near zero values.

\begin{figure}[htp!]
\centering
\includegraphics[width=\linewidth]{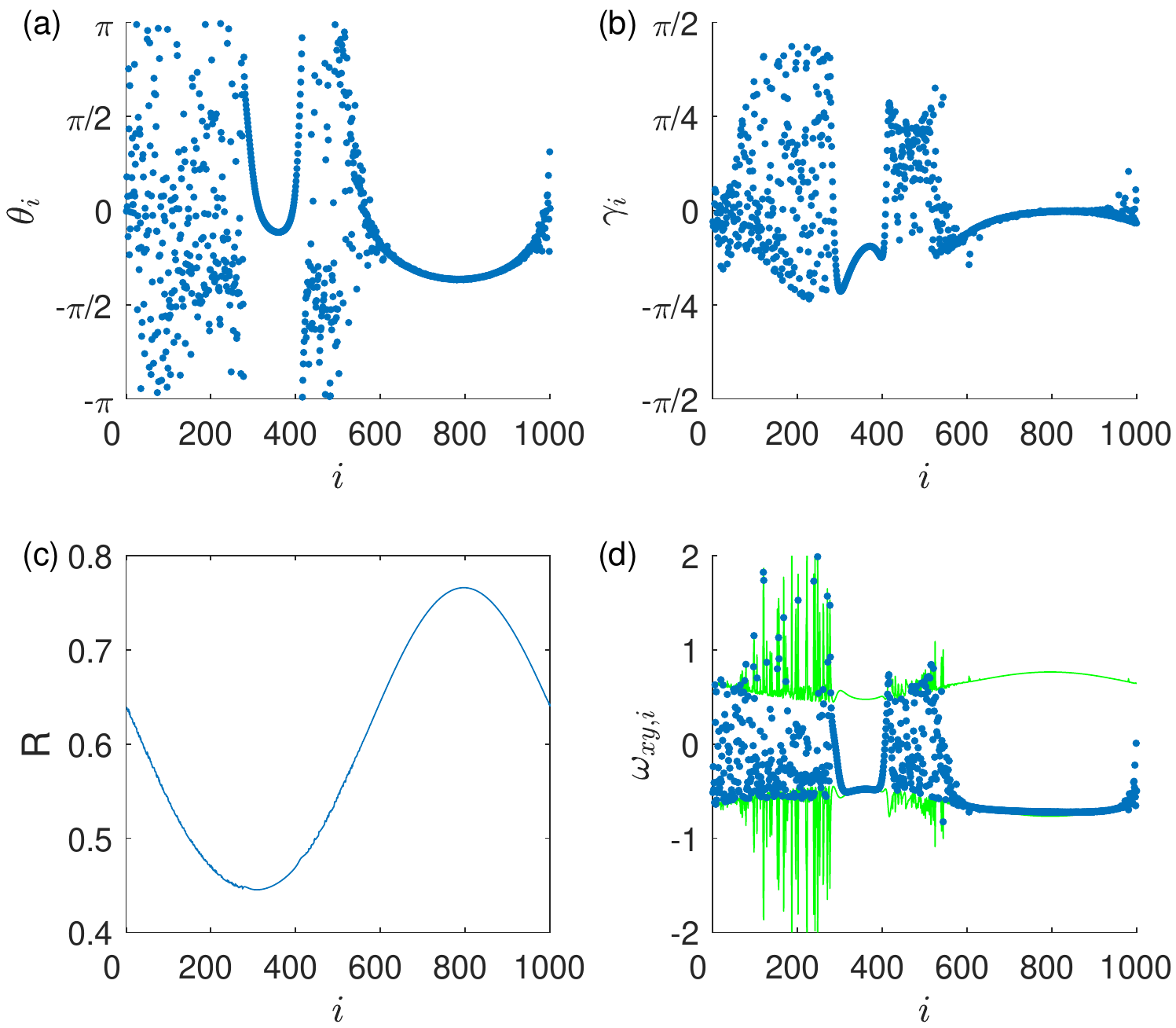}
\caption{ {\em Snapshots of the same system in Fig.~\ref{fig:stage12} 
in its third stage}. (a) Snapshot of the longitudinal angles $\theta_i$ of 
the oscillators. A new coherent region is formed at roughly $600<i<800$,
which is larger than the one in $300<i<400$ formed during the first 
stage shown in Figs.~\ref{fig:stage12}(a1,b1). The system is now in a 
multi-chimera state with several coherent and incoherent regions. (b) Snapshot 
of the latitudinal angles $\gamma_i$ of the oscillators. After the appearance 
of the new coherent region, the values of $\gamma_i$ of these oscillators 
in this region are close to zero, in contrast to the second stage. The
angles $\gamma_i$ of the oscillators from the old coherent region deviate 
further away from zero than in the first and second stages, but they are 
still smaller than the angles of most oscillators in the incoherent regions. 
(c) Snapshot of the order parameter $R$. The peak is now at the center of 
the new coherent region. (d) Snapshots of the instantaneous angular velocity 
in the $(x,y)$-plane for all the oscillators: values of $\omega_{xy,i}$ (blue 
dots) and $\pm R_i\chi_i$ (green traces).}
\label{fig:stage3}
\end{figure}

In the second stage, while the $\theta$ components of the oscillators are
still in a chimera state, the $\gamma$ components of some oscillators are 
no longer close to zero and the average absolute value of $\gamma$ in the 
incoherent region becomes larger than that in the coherent region, as shown 
in Fig.~\ref{fig:stage12}(b2). Such high values of $\gamma$ make the value 
of $\chi$ high as well. As a result, in this stage many oscillators in the 
incoherent region have a larger average $R\chi$ value than those in the 
coherent region, as shown in Fig.~\ref{fig:stage12}(b3), a behavior that is
opposite to that in the first stage. 
Moreover, from Eq.~(\ref{eq:SphericalTheta}), we have that 
the instantaneous phase or angular velocity $d\theta/dt$ in the $(x,y)$-plane, 
$\omega_{xy}$, is bounded by $\pm R\chi$. A larger $R\chi$ value in the 
incoherent region can thus result in a larger absolute value of the phase 
velocity $|\omega_{xy}|$ than that in the coherent region, as shown in 
Fig.~\ref{fig:stage12}(b3). Note that, in the first stage, similar to the 
transient 2D chimera states with a long lifetime, the oscillators in the 
coherent region have the largest possible value of $|\omega_{xy}|$, as shown 
in Fig.~\ref{fig:stage12}(a3), making the oscillators in the incoherent 
region unable to synchronize with the coherent oscillators since the 
incoherent region does not have similarly large $R\chi$ values required for
large values of $|\omega_{xy}|$. However, in the second stage, this is no 
longer the case: oscillators in the incoherent region can have $|\omega_{xy}|$
values larger than those in the coherent region, driving the system away 
from the 2D-like structure.

\begin{figure}[htp!]
\centering
\includegraphics[width=\linewidth]{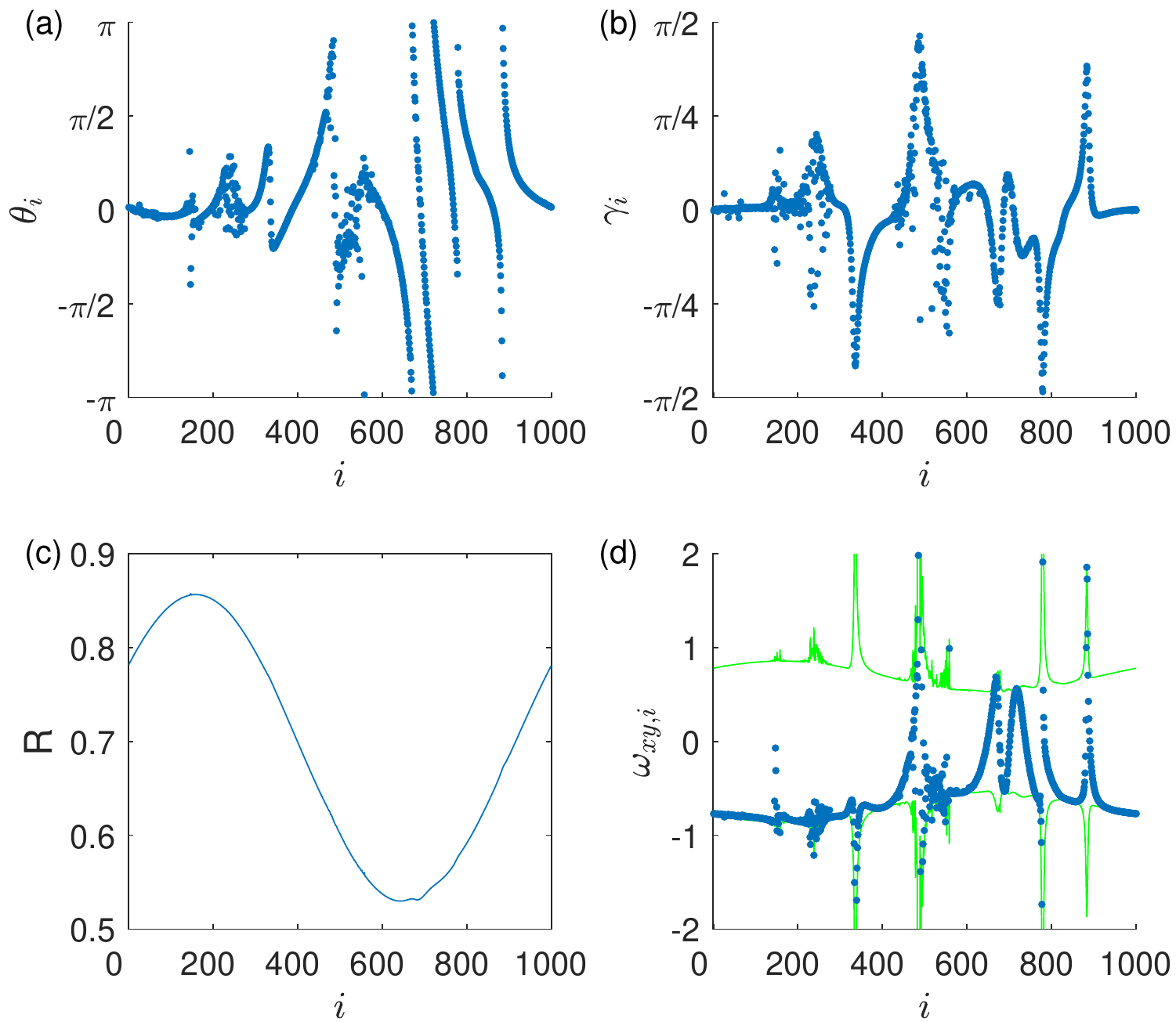}
\caption{ {\em Snapshots of the same system in Figs.~\ref{fig:stage12}
and \ref{fig:stage3} in the fourth stage}. (a,b) Snapshots of $\theta_i$ 
and $\gamma_i$. (c) Snapshot of the order parameter $R$. The peak is now at 
a new location away from its location in the third stage. (d) Snapshots of 
the instantaneous angular velocity in the $(x,y)$-plane of all the oscillators:
$\omega_{xy,i}$ (blue dots) and $\pm R_i\chi_i$ (green traces). The system is 
now more fragmental than in the third stage: the extensive coherent region 
from the third stage has now broken into several coherent and incoherent 
regions.}
\label{fig:stage4}
\end{figure}

As a consequence of large values of $R\chi$, in the third stage, a new and 
even wider coherent region emerges within the incoherent region, placing the
whole system in a transient multi-chimera state with two coherent regions: 
one is formed at the beginning of the first stage and the other newly appeared 
one emerging inside the incoherent region, as shown in Fig.~\ref{fig:stage3}. 
In the third stage, the region used to be the most incoherent becomes now the 
most coherent with the largest value of $R$ and a larger size than the original
coherent region. The $R$ value of the original coherent region is now close to 
its minimum value. These observations suggest a negative feedback mechanism. 
In particular, incoherence in the first stage results in small $R_i$ values 
and thus small $R_i\chi_i$ values as well, making $\gamma_i$ deviate away 
from zero, which in turn results in a high value of $R_i\chi_i$ that leads to 
incoherence. As a result, coherence in the system has been undermined, 
leading to near zero average value of $\gamma_i$, as the oscillators have 
evenly distributed positive and negative $\gamma_i$ values. This leads to a 
small average value of $R_i\chi_i$, making it difficult for the oscillators 
to remain coherent. While this mechanism does not turn coherence into 
incoherence as effectively as a large value of $R_i\chi_i$ turns incoherence 
into coherence, a fraction of the oscillators still have a large average 
$\gamma$ value. The difference drives the whole system towards global 
coherence.

\begin{figure}[htp!]
\centering
\includegraphics[width=\linewidth]{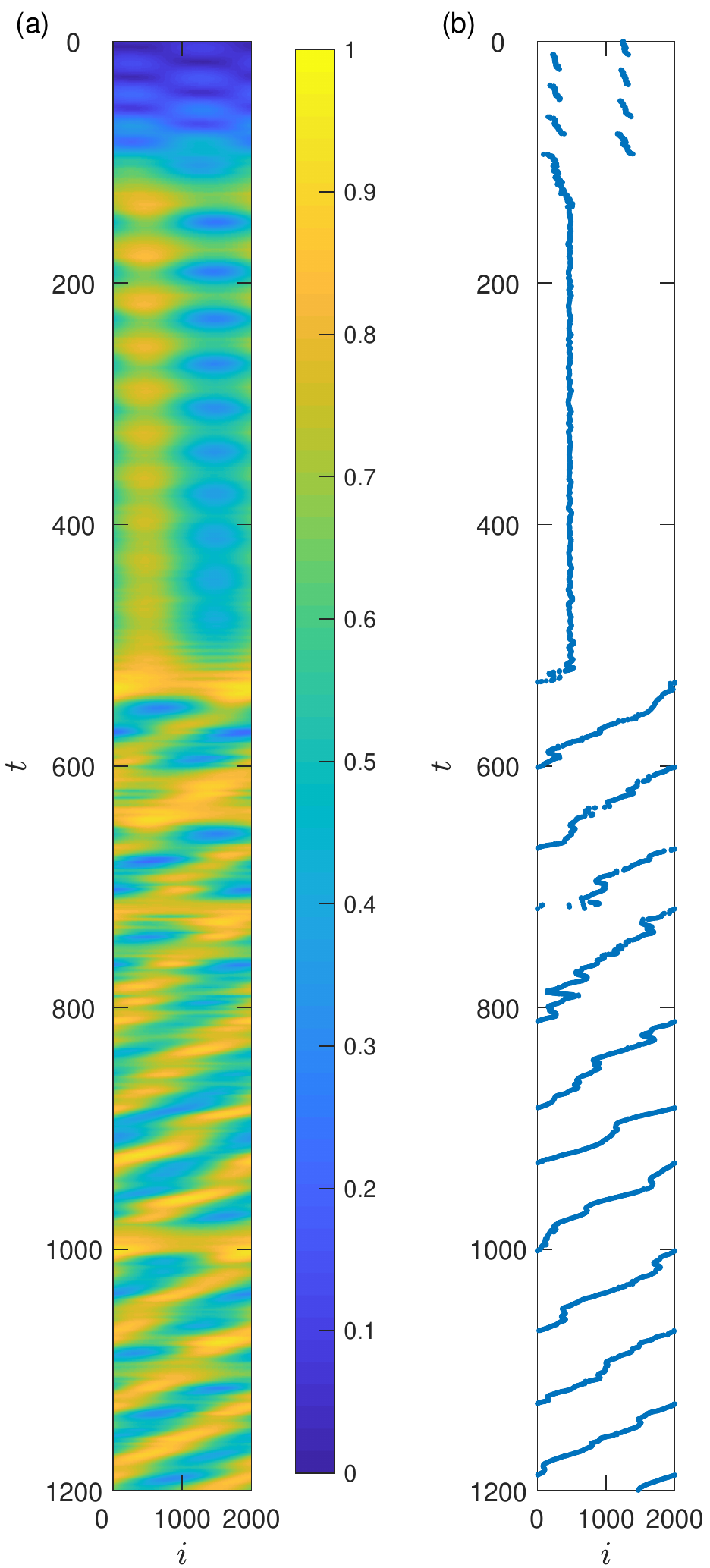}
\caption{ {\em Spatial distribution of the order parameter of the oscillators}.
(a) Value of the order parameter $R$ of all the oscillators.
(b) The location of the oscillator with the largest value of $R$ among all the
oscillators versus time, which is constantly traveling across the system
after the collapse of the 2D-like chimera state at $t=540.9$.}
\label{fig:stageR}
\end{figure}

In the fourth stage, a process similar to that in the third stage occurs,
where large coherent and incoherent regions break into smaller subregions
of coherence and incoherence. An example is shown in Fig.~\ref{fig:stage4}. 
The system becomes fragmental without any recognizable pattern. There are
two major sources of randomness. Firstly, oscillators in the same coherent 
region are phase-locked in the $\theta$ component, but the locked phase 
velocities are different from region to region. Because these coherent 
regions have different average values of $R$ and $R\chi$. Secondly, the 
positions of the newly formed regions are sensitive to the random initial 
condition, rendering random sizes of the regions.

In the final stage, the system is close to a globally coherent state, but can 
never actually reach it. The global state is dynamical, with the emergence and 
disappearance of many fragmental coherent regions driven by the negative 
feedback mechanism discussed above. Globally, $g_0$ reaches a relatively 
stable value with fluctuations, as shown in Fig.~\ref{fig:TCS}(b). Several 
snapshots of this stage are shown in Fig.~\ref{fig:snapshot}(b1-b4).

\begin{figure}[htp!]
\centering
\includegraphics[width=\linewidth]{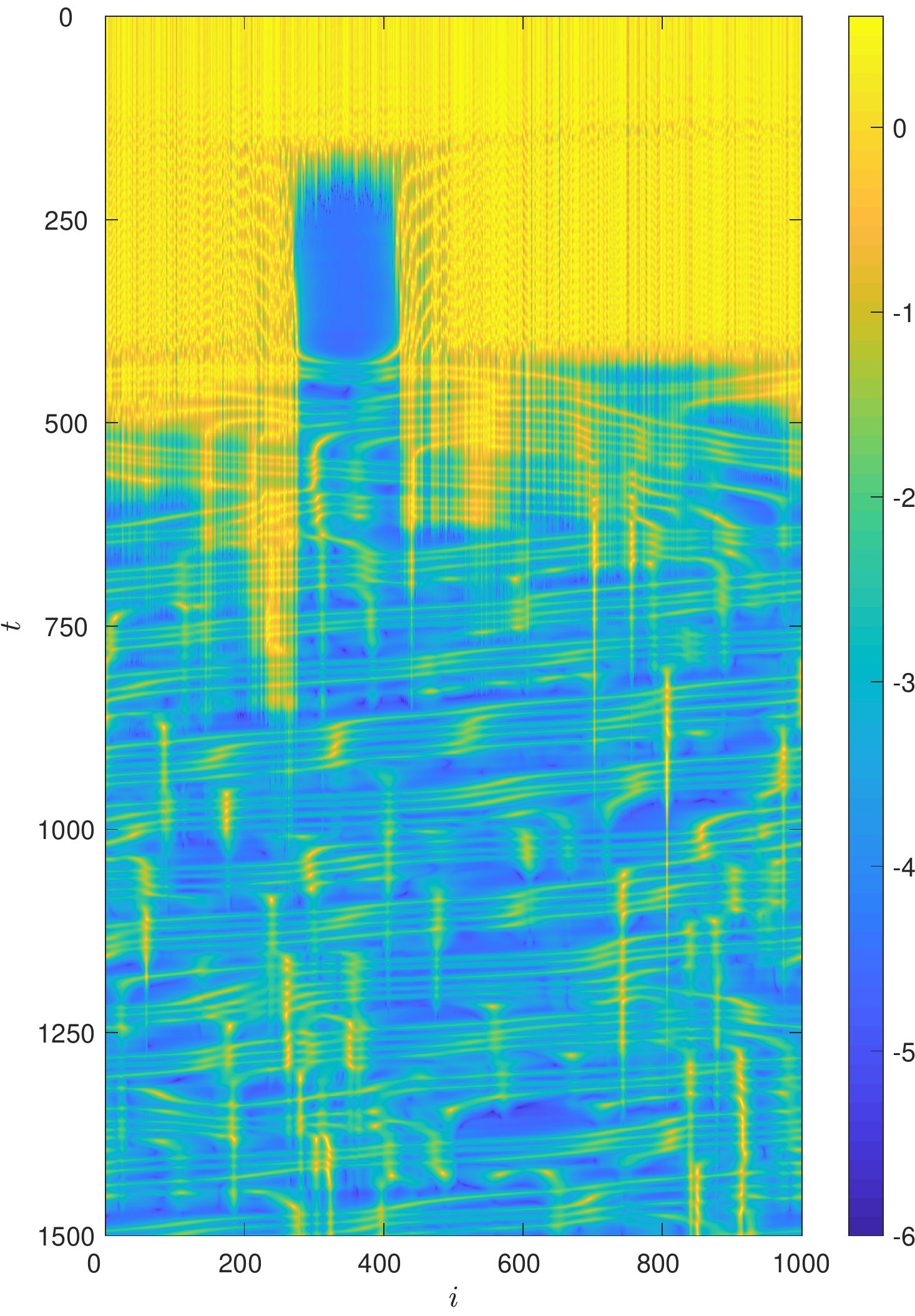}
\caption{ {\em Temporal evolution of $\log_{10}{D_i}$}. 
The regions with $\log_{10}{D_i}<-2$ are classified as coherent regions.
Other regions are classified as incoherent regions. Note that there is 
coexistence of one coherent and another incoherent regions in the system in 
the time interval $150 \alt t \alt 400$, indicating a transient chimera state.}
\label{fig:stageD}
\end{figure}

Figures~\ref{fig:stageR} and \ref{fig:stageD} present additional evidence 
to support our understanding of the fragility of the chimera state. In 
particular, Fig.~\ref{fig:stageR}(a) shows the location of the oscillators 
with the largest value of the order parameter $R$ among all other oscillators 
at a time instant. Because of the choice of a non-local/non-global coupling 
function $G(i-j)$ and the order parameter defined as 
$\boldsymbol{\rho}_i=N^{-1}\sum_{j=1}^NG (i-j)\boldsymbol{\sigma}_j$, the
position at which $R$ reaches the maximum value gives information about 
the center of the non-local/non-global coherent region. Prior to the collapse
of the 2D-like chimera state, this position remains at the center of the 
coherent region. However, after the collapse, the position constantly rotates 
among different oscillators. As shown in Fig.~\ref{fig:stageD}, the coherent 
regions with a low value of $\log_{10}{D_i}$, denoted by the blue color, 
have short lifetime in the stages after the collapse of the chimera state. 
During those stages, the values of $\log_{10}{D_i}$ of different regions 
in the system oscillate. This also agrees with our understanding that 
a high value of $\gamma$ would decrease soon due to the negative feedback 
mechanism.

The results studied so far are for the case where the values of the coupling
parameters $A$ and $\alpha$ are away from the boundary of the basin of the 
chimera states. As this boundary is approached, another type of final states
after the collapse of the transient chimera states appears, which is the 
global synchronization state. The fraction of the initial states leading 
to this synchronous state increases toward one. (Here we exclude the cases 
where no chimera state ever appears and the system directly goes to global
synchronization, and focus on the cases where there was a chimera state.)
The origin of this alternative final state can be understood as follows: 
near the basin boundary, the relative size of the coherent region associated 
with the chimera state becomes larger. As explained, in the third stage a
second and even larger coherent region will emerge in the middle of the
incoherent region. Since the first coherent region has already become 
relatively large (e.g., about half the size of the whole system), the
final state contains a large coherent region as in the third stage.

\begin{figure}[htp!]
\centering
\includegraphics[width=\linewidth]{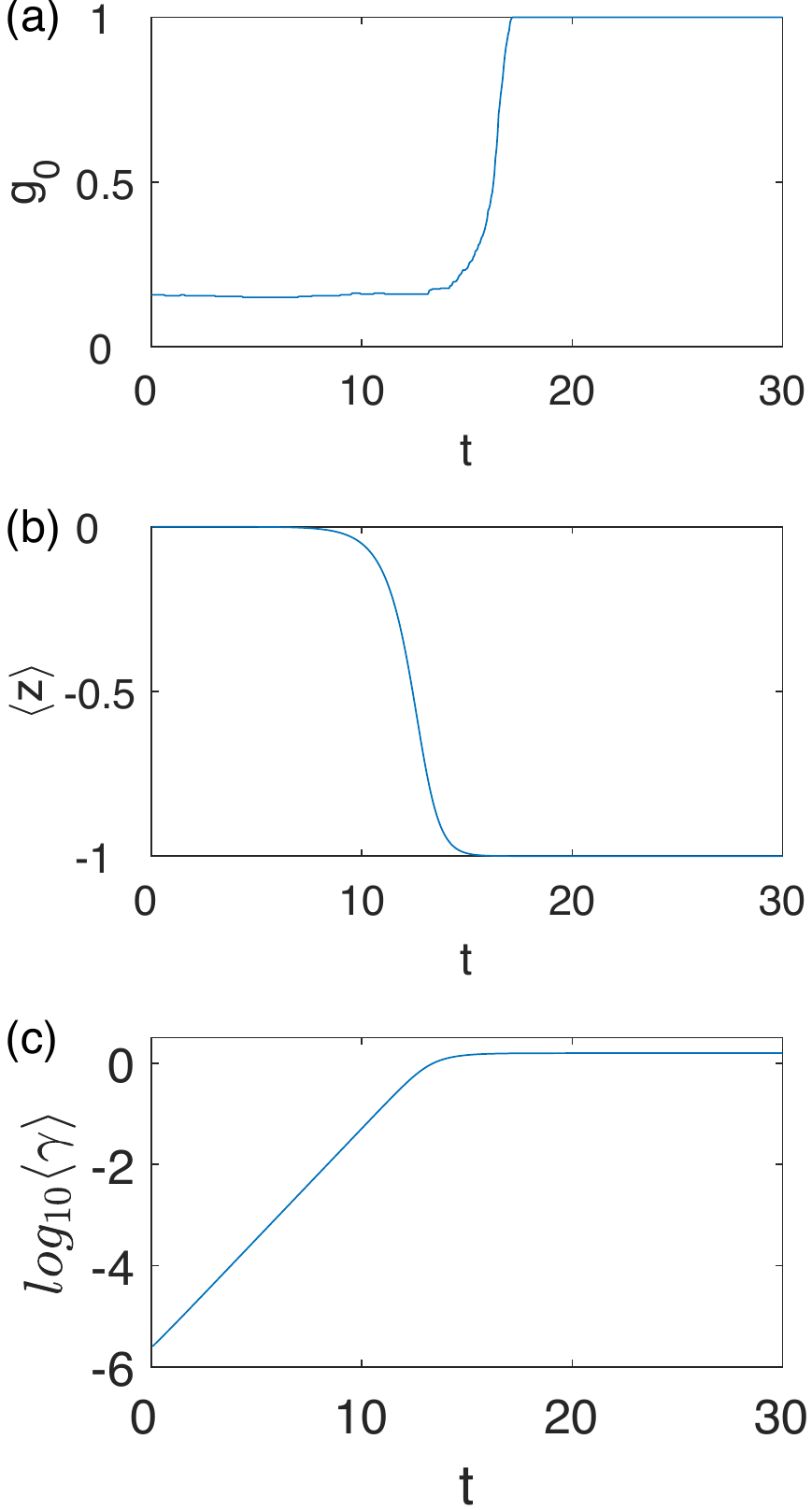}
\caption{ {\em Transient chimera state in 3D systems with $T_{33} = +1$}.
(a) Time evolution of $g_0$ with initial perturbation 
$\delta=10^{-3}$. The transient time before the system collapses into a global 
synchronized state (where $g_0=1$) is one order of magnitude shorter than that 
for the case of $T_{33} = -1$ for the same value of $\delta$ 
(c.f., Fig.~\ref{fig:tvsgammad}). (b) Time evolution of the average $z$ 
component (in Cartesian coordinates) of all the oscillators, which converges
rapidly to $-1$, indicating that all phase vectors $\sigma_i$ are concentrating
at the ``south pole'' of the unit sphere $S^2$ after a short transient.
(c) Logarithm of the time evolution of the average latitudinal angle of all 
oscillators. The approximately linear growth of the logarithm indicates 
an exponential growth of the average latitudinal angle during the transient.
System parameter values are $N=400$, $\theta_d=0.001$, $\alpha=\pi/2-0.05$, 
and $A=0.995$.}
\label{fig:det1}
\end{figure}

\section{3D systems with $T_{33} = +1$ and 4D systems} \label{sec:T33And4D}

We see from Sec.~\ref{sec:Model} that there is another possible case in 3D 
where $T_{33} = +1$. A difficulty is that, for $T_{33} = +1$,
any 2D structure is physically or computationally not observable because an
infinitesimal perturbation (e.g., on the order of the computer round off
error) is sufficient to destroy the 2D structure even before the emergency
of any transient chimera state. To overcome this difficulty, we first
generate chimera states in a purely 2D system, and then perturb these 2D 
chimera configurations and use them as the initial states for the 
$T_{33} = +1$ systems. The systems are then initially in chimera states.
With this initial chimera state, the oscillators still quickly synchronize to 
the fixed points of the transformation $\boldsymbol{{\rm T}}$: $(0,0,\pm 1)$ 
(in Cartesian coordinates), as shown in Fig.~\ref{fig:det1}. Since the 
longitudinal angles at the fixed points are zero, the phase lag plays no role 
in the dynamics. In the high-dimensional Kuramoto model with no phase 
lag, all the oscillators have the same natural frequency, making the 
synchronized state stable~\cite{CGO:2019}. 

\begin{figure}[htp!]
\centering
\includegraphics[width=\linewidth]{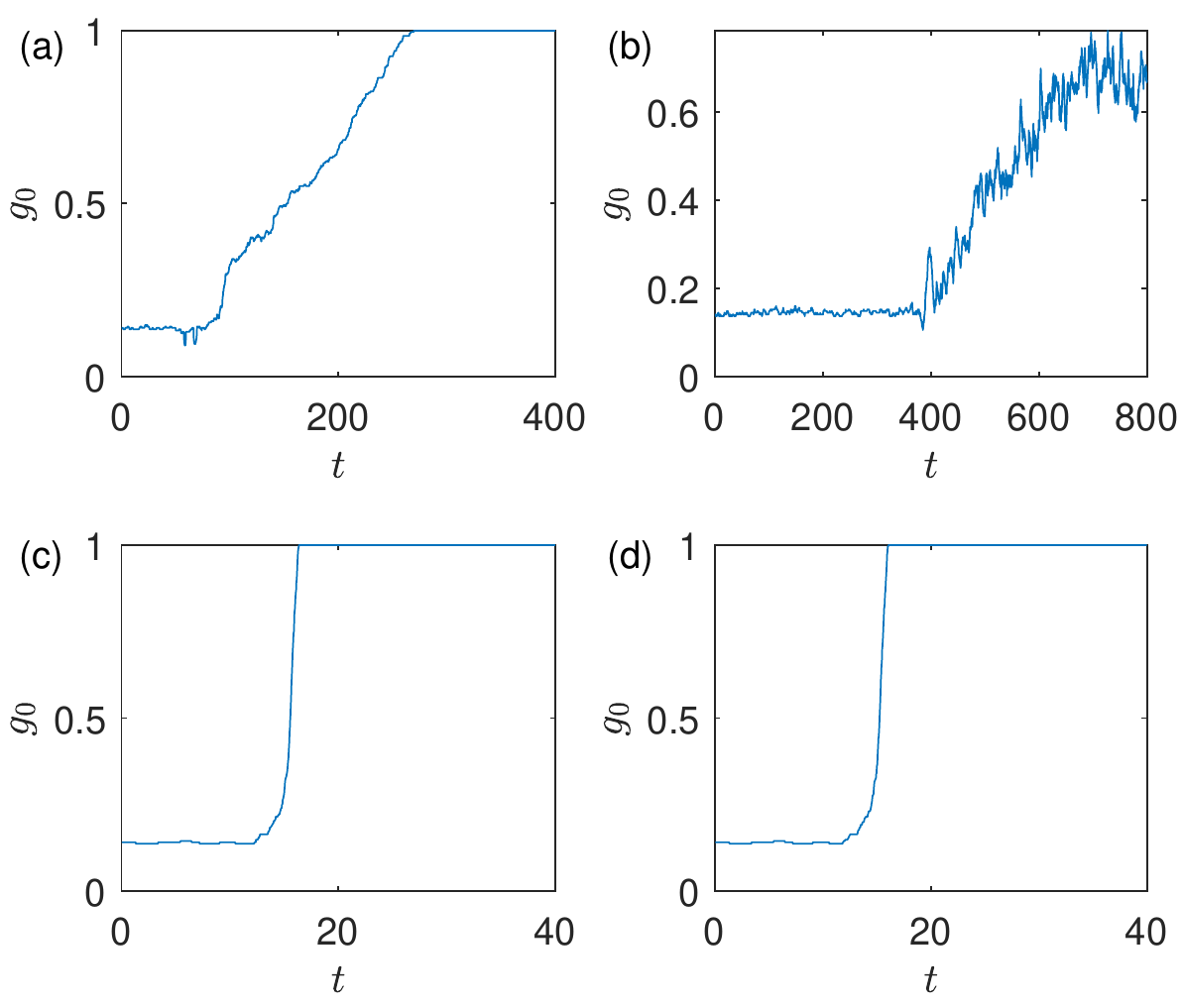}
\caption{ {\em Transient chimera states in 4D}. Shown is the time evolution
of $g_0$ in four different types of 4D systems: (a) $k=2$ and $m=0$,
(b) $k=1$ and $m=2$, (c) $k=1$ and $m=1$, (d) $k=1$ and $m=0$. 
In all four cases, the double rotation matrix $\boldsymbol{{\rm T}}$ has 
$\alpha_1=\pi/2-0.05$ and $\alpha_2=\pi/2-0.06$. All four systems start from
the same 2D chimera state as the initial conditions, with a perturbation
of $\phi_{2,d}=\phi_{3,d}=0.001$ in the second and third angular coordinates
in the 4D spherical coordinate system. The values of other system parameters
are $N=256$ and $A=0.995$. In all cases, the initial 2D chimera state can
last only for a short time before being destroyed.}
\label{fig:4D}
\end{figure}

Insights into this behavior can be gained by analyzing the linearized 
approximation of Eq.~(3), as the values of $\gamma_i$ and $\Gamma_i$ 
are typically small before the collapse:
\begin{equation} \label{eq:SphericalGammaLinearDet1}
\frac{d\gamma_i}{dt}=R_i\times \left[-cos\left(\theta_i-\Theta_i+\alpha\right)\gamma_i+\Gamma_i\right],
\end{equation}
where $R_i$ is positive. The quantity $\left(\theta_i-\Theta_i+\alpha\right)$
exhibits rapid oscillations in the interval $[-\pi,\pi)$ without an apparent
pattern, so the quantity $-\cos\left(\theta_i-\Theta_i+\alpha\right)$
oscillates within the interval $[-1,1]$ approximately randomly. The sign of
the factor $-R_i\times \cos\left(\theta_i-\Theta_i+\alpha\right)\gamma_i$
thus changes rapidly, while the sign of $R_i$ remains positive. As a result,
the $R_i \Gamma_i$ component plays a major role in the dynamics, making the
absolute value of $\gamma_i$ increase exponentially. This agrees with the
simulation result, as shown in Fig.~\ref{fig:det1}(c), where we observe
that the average $\gamma$ value indeed grows exponentially in time.

We see from Sec.~\ref{sec:Model} that four different forms of the 
matrix $\boldsymbol{{\rm T}}$ can arise in 4D. We encounter the
similar difficulty in all these four forms as in the $T_{33} = +1$ case in 3D:
an infinitesimal perturbation on the order of the computer round off
error is sufficient to destroy the 2D structure before the emergency
of any transient chimera state. Thus in all the four different cases
we let the system begin with 2D chimera states as for the $T_{33} = +1$ case 
in 3D. Figure~\ref{fig:4D} shows the representative results on the dynamical 
evolution of a 4D system from the initial chimera state. In all cases, the 
chimera state can last for a relatively short time only, indicating that, in 
4D, chimera states (even of an appreciable transient lifetime) are unlikely 
to occur.

\section{Discussion} \label{sec:discussion}

To summarize, our finding of an extreme type of fragility of chimera states 
against dimension-augmenting perturbations beyond 2D, quantitatively 
characterized by a logarithmic scaling law, has implications with respect to 
observability. We also note that, in 2D, when the number $N$ of oscillators 
is finite, chimera states are long transients~\cite{WO:2011} and their
average lifetime increases exponentially with $N$. However, in our case, 
the transient lifetime increases logarithmically with $N$. Indeed, we have 
never observed long transients in higher dimensions, even when there are 
thousands of oscillators in the system.

Besides the characteristically distinct scaling laws of the average transient
lifetime, there are other differences between the 2D and dimension-augmenting 
transient chimera states. In 2D, the distribution of the transient lifetime
from random initial conditions is exponential. However, in our model, the
distribution is approximately Gaussian, with a small standard deviation (less 
than 10\% of the mean value, as shown in 
Figs.~\ref{fig:tvsgammad}-\ref{fig:6}). This means that the transient lifetime 
is much more closely clustered near the mean value in our model than in the 2D 
case. In general, when the system is in a particular transient chimera state, 
it is difficult to predict when it will collapse into 2D, as the transient 
states preceding the collapse exhibit a similar pattern. However, in our model,
we can use a variable such as the spatial 
variance of $\gamma$ [as in Fig.~\ref{fig:TCS}(c)] as an indicator to predict 
any possible collapse, as it tends to increase monotonously towards a threshold
value. In 2D, the transient chimera states are reported to belong to type-II 
supertransients~\cite{WO:2011} by the criterion in Ref.~\cite{CK:1988}, 
which have the following features: an exponential scaling law of the transient 
lifetime with respect to system size, the exponential distribution of the 
transient lifetime, and stationarity of the patterns. Transient chimera states 
in 2D have all these features, but none can be found in our 3D or 4D systems. 
This further indicates the characteristically different nature of the 
transient chimera states in dimension-augmented systems from those in 2D.

There were previous studies on transient chimera states with respect to 
changes in the inertia of the oscillators~\cite{Olmi:2015,Laing:2019} where,
in the model studied, a change in the inertia from a non-zero value to zero
can abruptly alter the dynamical behavior of the system, making the chimera
state more unstable. Our results demonstrate that a dimension-augmenting 
perturbation can also make the chimera state extremely unstable in the 
sense of the scaling law uncovered. However, our model is the first-order 
Kuramoto system with 3D phase oscillators and the model previously 
studied~\cite{Olmi:2015,Laing:2019} is a second-order phase Kuramoto system 
with inertia. The system settings are thus quite different. Further, the 
results are quite different as well: our main result is the logarithmic 
scaling law of the average transient lifetime of the chimera state with respect 
to both the perturbation magnitude, which holds regardless of the system 
size. In the previous studies~\cite{Olmi:2015,Laing:2019}, no such scaling 
law was reported; instead, the scaling law of the intermittent chaotic 
chimeras lifetime therein was found to be algebraic.

The chimera states can be long lasting when the dynamics of the oscillators are 
strictly equivalent to those of 2D rotors. Any deviation in the rotation 
dynamics from 2D will make a chimera state transient, with an extraordinarily 
short lifetime in the sense of the scaling law uncovered. However, there
are situations where chimera states with a short lifetime may still be      
physically meaningful and observable. For example, for a highly dynamic 
system whose state changes constantly with time, the intrinsic time scale is 
short. In this case, if the time scale is shorter than the average transient 
lifetime of the chimera states, they are still physically meaningful and 
observable.

A recent analysis of synchronization in high-dimensional Kuramoto 
models~\cite{CGO:2019b} revealed the existence of an invariant manifold 
of state distribution in the thermodynamic limit $N\to\infty$, which is an 
extension of the previous work on the 2D Kuramoto model~\cite{OA:2008,OA:2009}.
This manifold is attracting in 2D~\cite{OA:2009} and is likely to be 
attracting in high dimensions as well. However, chimera states, even in 2D, 
do not live on the invariant manifold. In higher dimensions, the manifold is 
circularly symmetric about some axis and the density of the distribution 
decreases monotonously from one axis pole to another. If the manifold is 
attracting, a 2D chimera state would naturally collapse and approach the 
attracting manifold, even when the system size is large. In higher dimensions, 
the picture is less clear as to whether the manifold is attracting. 
Identifying the existence of some invariant manifold and determining its 
stability may provide an avenue to study collective dynamics in 
high-dimensional Kuramoto models.

We remark that global synchronization in the high-dimensional Kuramoto model 
has been treated~\cite{CGO:2019,CGO:2019b}, where it was shown that 
synchronization is stable under small perturbations. Global and cluster 
synchronization states in the high-dimensional Kuramoto model and their 
stability are different from our problem of the effect of dimension-augmenting 
perturbations on chimera states. In our setting, to generate chimera states, 
there are phase lags among the oscillators but their natural frequencies are 
identical. The common, natural rotation of the oscillators can then be 
excluded from consideration through the reference frame rotating at the same 
frequency. This should be contrasted to the setting of global and cluster 
synchronization, where the natural frequencies of the oscillators are 
different and follow a certain distribution and this heterogeneity plays an 
important role in synchronization. For example, for even dimensional models 
the criteria of whether an oscillator belongs to the entrained population or 
the drifting population depends on the value of the frequency of the natural 
rotation of the oscillator.

\section*{Acknowledgment}

We would like to acknowledge support from the Vannevar Bush Faculty
Fellowship program sponsored by the Basic Research Office of the Assistant
Secretary of Defense for Research and Engineering and funded by the Office
of Naval Research through Grant No.~N00014-16-1-2828.

\appendix

\section{Measure for detecting transient chimera states and determination
of their lifetime} \label{sec:SMDistChimeraState}

For all the simulations we use the fourth-order Runge-Kutta method to
integrate the coupled networked system with time step $dt=0.005$. In
particular, we integrate the system with the order parameter defined as
$\boldsymbol{\rho}_i=N^{-1}\sum_{j=1}^NG (i-j)\boldsymbol{\sigma}_j$.
When the local order parameter of oscillator $i$ includes a small
self-coupling component, the average lifetime of the chimera state is
longer than that in systems without such self-coupling. However, inclusion
of the self-coupling term has little effect on the scaling law between
the average chimera time and the magnitude of the perturbation.

To ascertain the existence of a transient chimera state, we use the measure
$g_0$ introduced in Ref.~\cite{KHSKK:2016}, which is the relative size of
the coherent region. To calculate $g_0$, another quantity $D_i$ is needed,
which is the spatial Laplacian of $\boldsymbol{\sigma}$ at oscillator $i$
that characterizes the instant local degree of distortion of its dynamical
variables:
\begin{eqnarray} \nonumber
D_i&=&[\left( x_{i+1}-2x_i+x_{i-1}\right)^2+\left( y_{i+1}-2y_i+y_{i-1}\right)^2\\ 
&+&\left( z_{i+1}-z_i+z_{i-1}\right)^2]^{1/2},
\end{eqnarray}
where $x_i$, $y_i$ and $z_i$ are the Cartesian coordinates of the 3D
state vector of oscillator $i$. If the value of $D_i$ is smaller than a
certain threshold, oscillator $i$ is regarded as being within the coherent
region; otherwise it is in an incoherent region. The value of $g_0$ at time
$t$ is calculated by counting the number of oscillators with $D_i$ smaller
than the threshold at this time, normalized by the total number $N$ of
oscillators. In our study, we choose the threshold value of $D_i$ to be
$0.04$, which is approximately one hundredth of the upper bound of $D_i$.

When the system exhibits a chimera state, coherent and incoherent regions
coexist, so we have $0\le g_0 \le 1$. Furthermore, $g_0$ will be plateaued at a
certain value with small fluctuations about it~\cite{KHSKK:2016}. This
provides a convenient and effective way to detect the transient chimera state.
Especially, when the value of $g_0$ reaches a relatively flat plateau,
the chimera state begins. The transient chimera state ends when the value
of $g_0$ begins to deviate from the plateaued value.

Our algorithm to find the exact starting time and ending time of a transient
chimera state can be described, as follows.
\begin{enumerate}
\item \vspace*{-0.1in}
Roughly choose a interval of the value of $g_0$ which include the values
of $g_0$ in the transient chimera states. With a certain set of system 
parameters, the value of $g_0$ in the chimera states are roughly similar.
This interval does not need to be very accurate.
\item \vspace*{-0.1in}
Determine the time interval in which the value of $g_0$ is in the interval
chosen in step 1.
\item \vspace*{-0.1in}
Calculate the average $g_0$ value in this time interval.
\item \vspace*{-0.1in}
Measure the starting time of chimera state as the first time $g_0$ reaches
the average value minus a small threshold value. We choose the threshold to
be $0.05$ in this paper.
\item \vspace*{-0.1in}
Determine the ending time of the chimera state as the last time when the
value of $g_0$ is smaller than the same average adding the threshold value,
before the value of $g_0$ ever grows beyond the average plus two times
the threshold value.
\item \vspace*{-0.1in}
The starting and ending time so determined can be inaccurate because of the
chosen initial time interval from step 2. To increase the accuracy, we set the 
period between the starting and ending time as the new initial time interval.
\item \vspace*{-0.1in}
Repeat steps 3-6 for 5 times to make the results converge.
\end{enumerate}
We determine the ending point this way because $g_0$ sometimes fluctuates
in a chimera state. The fluctuations arise even in purely 2D chimera states
and can be relatively large. We choose some appropriate threshold value
(e.g., $0.05$) to distinguish the cases when the system is in a chimera
state and when it has left the state.


\begin{thebibliography}{77}%
\makeatletter
\providecommand \@ifxundefined [1]{%
 \@ifx{#1\undefined}
}%
\providecommand \@ifnum [1]{%
 \ifnum #1\expandafter \@firstoftwo
 \else \expandafter \@secondoftwo
 \fi
}%
\providecommand \@ifx [1]{%
 \ifx #1\expandafter \@firstoftwo
 \else \expandafter \@secondoftwo
 \fi
}%
\providecommand \natexlab [1]{#1}%
\providecommand \enquote  [1]{``#1''}%
\providecommand \bibnamefont  [1]{#1}%
\providecommand \bibfnamefont [1]{#1}%
\providecommand \citenamefont [1]{#1}%
\providecommand \href@noop [0]{\@secondoftwo}%
\providecommand \href [0]{\begingroup \@sanitize@url \@href}%
\providecommand \@href[1]{\@@startlink{#1}\@@href}%
\providecommand \@@href[1]{\endgroup#1\@@endlink}%
\providecommand \@sanitize@url [0]{\catcode `\\12\catcode `\$12\catcode
  `\&12\catcode `\#12\catcode `\^12\catcode `\_12\catcode `\%12\relax}%
\providecommand \@@startlink[1]{}%
\providecommand \@@endlink[0]{}%
\providecommand \url  [0]{\begingroup\@sanitize@url \@url }%
\providecommand \@url [1]{\endgroup\@href {#1}{\urlprefix }}%
\providecommand \urlprefix  [0]{URL }%
\providecommand \Eprint [0]{\href }%
\providecommand \doibase [0]{http://dx.doi.org/}%
\providecommand \selectlanguage [0]{\@gobble}%
\providecommand \bibinfo  [0]{\@secondoftwo}%
\providecommand \bibfield  [0]{\@secondoftwo}%
\providecommand \translation [1]{[#1]}%
\providecommand \BibitemOpen [0]{}%
\providecommand \bibitemStop [0]{}%
\providecommand \bibitemNoStop [0]{.\EOS\space}%
\providecommand \EOS [0]{\spacefactor3000\relax}%
\providecommand \BibitemShut  [1]{\csname bibitem#1\endcsname}%
\let\auto@bib@innerbib\@empty
\bibitem [{\citenamefont {Umberger}\ \emph {et~al.}(1989)\citenamefont
  {Umberger}, \citenamefont {Grebogi}, \citenamefont {Ott},\ and\ \citenamefont
  {Afeyan}}]{UGOA:1989}%
  \BibitemOpen
  \bibfield  {author} {\bibinfo {author} {\bibfnamefont {D.~K.}\ \bibnamefont
  {Umberger}}, \bibinfo {author} {\bibfnamefont {C.}~\bibnamefont {Grebogi}},
  \bibinfo {author} {\bibfnamefont {E.}~\bibnamefont {Ott}}, \ and\ \bibinfo
  {author} {\bibfnamefont {B.}~\bibnamefont {Afeyan}},\ }\bibfield  {title}
  {\enquote {\bibinfo {title} {Spatiotemporal dynamics in a dispersively
  coupled chain of nonlinear oscillators},}\ }\href@noop {} {\bibfield
  {journal} {\bibinfo  {journal} {Phys. Rev. A}\ }\textbf {\bibinfo {volume}
  {39}},\ \bibinfo {pages} {4835} (\bibinfo {year} {1989})}\BibitemShut
  {NoStop}%
\bibitem [{\citenamefont {Kuramoto}\ and\ \citenamefont
  {Battogtokh}(2002)}]{KB:2002}%
  \BibitemOpen
  \bibfield  {author} {\bibinfo {author} {\bibfnamefont {Y.}~\bibnamefont
  {Kuramoto}}\ and\ \bibinfo {author} {\bibfnamefont {D.}~\bibnamefont
  {Battogtokh}},\ }\bibfield  {title} {\enquote {\bibinfo {title} {Coexistence
  of coherence and incoherence in nonlocally coupled phase oscillators},}\
  }\href@noop {} {\bibfield  {journal} {\bibinfo  {journal} {Nonlin. Phenom.
  Complex Syst.}\ }\textbf {\bibinfo {volume} {5}},\ \bibinfo {pages} {380}
  (\bibinfo {year} {2002})}\BibitemShut {NoStop}%
\bibitem [{\citenamefont {Abrams}\ and\ \citenamefont
  {Strogatz}(2004)}]{AS:2004}%
  \BibitemOpen
  \bibfield  {author} {\bibinfo {author} {\bibfnamefont {D.~M.}\ \bibnamefont
  {Abrams}}\ and\ \bibinfo {author} {\bibfnamefont {S.~H.}\ \bibnamefont
  {Strogatz}},\ }\bibfield  {title} {\enquote {\bibinfo {title} {Chimera states
  for coupled oscillators},}\ }\href@noop {} {\bibfield  {journal} {\bibinfo
  {journal} {Phys. Rev. Lett.}\ }\textbf {\bibinfo {volume} {93}},\ \bibinfo
  {pages} {174102} (\bibinfo {year} {2004})}\BibitemShut {NoStop}%
\bibitem [{\citenamefont {Shima}\ and\ \citenamefont
  {Kuramoto}(2004)}]{SK:2004}%
  \BibitemOpen
  \bibfield  {author} {\bibinfo {author} {\bibfnamefont {S.-I.}\ \bibnamefont
  {Shima}}\ and\ \bibinfo {author} {\bibfnamefont {Y.}~\bibnamefont
  {Kuramoto}},\ }\bibfield  {title} {\enquote {\bibinfo {title} {Rotating
  spiral waves with phase-randomized core in nonlocally coupled oscillators},}\
  }\href@noop {} {\bibfield  {journal} {\bibinfo  {journal} {Phys. Rev. E}\
  }\textbf {\bibinfo {volume} {69}},\ \bibinfo {pages} {036213} (\bibinfo
  {year} {2004})}\BibitemShut {NoStop}%
\bibitem [{\citenamefont {Abrams}\ and\ \citenamefont
  {Strogatz}(2006)}]{AS:2006}%
  \BibitemOpen
  \bibfield  {author} {\bibinfo {author} {\bibfnamefont {D.~M.}\ \bibnamefont
  {Abrams}}\ and\ \bibinfo {author} {\bibfnamefont {S.~H.}\ \bibnamefont
  {Strogatz}},\ }\bibfield  {title} {\enquote {\bibinfo {title} {Chimera states
  in a ring of nonlocally coupled oscillators},}\ }\href@noop {} {\bibfield
  {journal} {\bibinfo  {journal} {Int. J. Bif. Chaos}\ }\textbf {\bibinfo
  {volume} {16}},\ \bibinfo {pages} {21} (\bibinfo {year} {2006})}\BibitemShut
  {NoStop}%
\bibitem [{\citenamefont {Abrams}\ \emph {et~al.}(2008)\citenamefont {Abrams},
  \citenamefont {Mirollo}, \citenamefont {Strogatz},\ and\ \citenamefont
  {Wiley}}]{AMSW:2008}%
  \BibitemOpen
  \bibfield  {author} {\bibinfo {author} {\bibfnamefont {D.~M.}\ \bibnamefont
  {Abrams}}, \bibinfo {author} {\bibfnamefont {R.}~\bibnamefont {Mirollo}},
  \bibinfo {author} {\bibfnamefont {S.~H.}\ \bibnamefont {Strogatz}}, \ and\
  \bibinfo {author} {\bibfnamefont {D.~A.}\ \bibnamefont {Wiley}},\ }\bibfield
  {title} {\enquote {\bibinfo {title} {Solvable model for chimera states of
  coupled oscillators},}\ }\href@noop {} {\bibfield  {journal} {\bibinfo
  {journal} {Phys. Rev. Lett.}\ }\textbf {\bibinfo {volume} {101}},\ \bibinfo
  {pages} {084103} (\bibinfo {year} {2008})}\BibitemShut {NoStop}%
\bibitem [{\citenamefont {Sethia}\ \emph {et~al.}(2008)\citenamefont {Sethia},
  \citenamefont {Sen},\ and\ \citenamefont {Atay}}]{SSA:2008}%
  \BibitemOpen
  \bibfield  {author} {\bibinfo {author} {\bibfnamefont {G.~C.}\ \bibnamefont
  {Sethia}}, \bibinfo {author} {\bibfnamefont {A.}~\bibnamefont {Sen}}, \ and\
  \bibinfo {author} {\bibfnamefont {F.~M.}\ \bibnamefont {Atay}},\ }\bibfield
  {title} {\enquote {\bibinfo {title} {Clustered chimera states in
  delay-coupled oscillator systems},}\ }\href@noop {} {\bibfield  {journal}
  {\bibinfo  {journal} {Phys. Rev. Lett.}\ }\textbf {\bibinfo {volume} {100}},\
  \bibinfo {pages} {144102} (\bibinfo {year} {2008})}\BibitemShut {NoStop}%
\bibitem [{\citenamefont {Laing}(2009)}]{Laing:2009a}%
  \BibitemOpen
  \bibfield  {author} {\bibinfo {author} {\bibfnamefont {C.~R.}\ \bibnamefont
  {Laing}},\ }\bibfield  {title} {\enquote {\bibinfo {title} {Chimera states in
  heterogeneous networks},}\ }\href@noop {} {\bibfield  {journal} {\bibinfo
  {journal} {Chaos}\ }\textbf {\bibinfo {volume} {19}},\ \bibinfo {pages}
  {013113} (\bibinfo {year} {2009})}\BibitemShut {NoStop}%
\bibitem [{\citenamefont {Sheeba}\ \emph {et~al.}(2009)\citenamefont {Sheeba},
  \citenamefont {Chandrasekar},\ and\ \citenamefont {Lakshmanan}}]{SCL:2009}%
  \BibitemOpen
  \bibfield  {author} {\bibinfo {author} {\bibfnamefont {J.~H.}\ \bibnamefont
  {Sheeba}}, \bibinfo {author} {\bibfnamefont {V.~K.}\ \bibnamefont
  {Chandrasekar}}, \ and\ \bibinfo {author} {\bibfnamefont {M.}~\bibnamefont
  {Lakshmanan}},\ }\bibfield  {title} {\enquote {\bibinfo {title} {Globally
  clustered chimera states in delay-coupled populations},}\ }\href@noop {}
  {\bibfield  {journal} {\bibinfo  {journal} {Phys. Rev. E}\ }\textbf {\bibinfo
  {volume} {79}},\ \bibinfo {pages} {055203} (\bibinfo {year}
  {2009})}\BibitemShut {NoStop}%
\bibitem [{\citenamefont {Martens}(2010)}]{Martens:2010}%
  \BibitemOpen
  \bibfield  {author} {\bibinfo {author} {\bibfnamefont {E.~A.}\ \bibnamefont
  {Martens}},\ }\bibfield  {title} {\enquote {\bibinfo {title} {Chimeras in a
  network of three oscillator populations with varying network topology},}\
  }\href@noop {} {\bibfield  {journal} {\bibinfo  {journal} {Chaos}\ }\textbf
  {\bibinfo {volume} {20}},\ \bibinfo {pages} {043122} (\bibinfo {year}
  {2010})}\BibitemShut {NoStop}%
\bibitem [{\citenamefont {Martens}\ \emph {et~al.}(2010)\citenamefont
  {Martens}, \citenamefont {Laing},\ and\ \citenamefont {Strogatz}}]{MLS:2010}%
  \BibitemOpen
  \bibfield  {author} {\bibinfo {author} {\bibfnamefont {E.~A.}\ \bibnamefont
  {Martens}}, \bibinfo {author} {\bibfnamefont {C.~R.}\ \bibnamefont {Laing}},
  \ and\ \bibinfo {author} {\bibfnamefont {S.~H.}\ \bibnamefont {Strogatz}},\
  }\bibfield  {title} {\enquote {\bibinfo {title} {Solvable model of spiral
  wave chimeras},}\ }\href@noop {} {\bibfield  {journal} {\bibinfo  {journal}
  {Phys. Rev. Lett.}\ }\textbf {\bibinfo {volume} {104}},\ \bibinfo {pages}
  {044101} (\bibinfo {year} {2010})}\BibitemShut {NoStop}%
\bibitem [{\citenamefont {Omel'chenko}\ \emph {et~al.}(2010)\citenamefont
  {Omel'chenko}, \citenamefont {Wolfrum},\ and\ \citenamefont
  {Maistrenko}}]{OWM:2010}%
  \BibitemOpen
  \bibfield  {author} {\bibinfo {author} {\bibfnamefont {O.~E.}\ \bibnamefont
  {Omel'chenko}}, \bibinfo {author} {\bibfnamefont {M.}~\bibnamefont
  {Wolfrum}}, \ and\ \bibinfo {author} {\bibfnamefont {Y.~L.}\ \bibnamefont
  {Maistrenko}},\ }\bibfield  {title} {\enquote {\bibinfo {title} {Chimera
  states as chaotic spatiotemporal patterns},}\ }\href@noop {} {\bibfield
  {journal} {\bibinfo  {journal} {Phys. Rev. E}\ }\textbf {\bibinfo {volume}
  {81}},\ \bibinfo {pages} {065201} (\bibinfo {year} {2010})}\BibitemShut
  {NoStop}%
\bibitem [{\citenamefont {Wolfrum}\ and\ \citenamefont
  {Omel'chenko}(2011)}]{WO:2011}%
  \BibitemOpen
  \bibfield  {author} {\bibinfo {author} {\bibfnamefont {M.}~\bibnamefont
  {Wolfrum}}\ and\ \bibinfo {author} {\bibfnamefont {O.~E.}\ \bibnamefont
  {Omel'chenko}},\ }\bibfield  {title} {\enquote {\bibinfo {title} {Chimera
  states are chaotic transients},}\ }\href@noop {} {\bibfield  {journal}
  {\bibinfo  {journal} {Phys. Rev. E}\ }\textbf {\bibinfo {volume} {84}},\
  \bibinfo {pages} {015201} (\bibinfo {year} {2011})}\BibitemShut {NoStop}%
\bibitem [{\citenamefont {Wolfrum}\ \emph {et~al.}(2011)\citenamefont
  {Wolfrum}, \citenamefont {Omel'chenko}, \citenamefont {Yanchuk},\ and\
  \citenamefont {Maistrenko}}]{WOYM:2011}%
  \BibitemOpen
  \bibfield  {author} {\bibinfo {author} {\bibfnamefont {M.}~\bibnamefont
  {Wolfrum}}, \bibinfo {author} {\bibfnamefont {O.~E.}\ \bibnamefont
  {Omel'chenko}}, \bibinfo {author} {\bibfnamefont {S.}~\bibnamefont
  {Yanchuk}}, \ and\ \bibinfo {author} {\bibfnamefont {Y.~L.}\ \bibnamefont
  {Maistrenko}},\ }\bibfield  {title} {\enquote {\bibinfo {title} {Spectral
  properties of chimera states},}\ }\href@noop {} {\bibfield  {journal}
  {\bibinfo  {journal} {Chaos}\ }\textbf {\bibinfo {volume} {21}},\ \bibinfo
  {pages} {013112} (\bibinfo {year} {2011})}\BibitemShut {NoStop}%
\bibitem [{\citenamefont {Omelchenko}\ \emph {et~al.}(2011)\citenamefont
  {Omelchenko}, \citenamefont {Maistrenko}, \citenamefont {H{\"o}vel},\ and\
  \citenamefont {Sch{\"o}ll}}]{OMHS:2011}%
  \BibitemOpen
  \bibfield  {author} {\bibinfo {author} {\bibfnamefont {I.}~\bibnamefont
  {Omelchenko}}, \bibinfo {author} {\bibfnamefont {Y.}~\bibnamefont
  {Maistrenko}}, \bibinfo {author} {\bibfnamefont {P.}~\bibnamefont
  {H{\"o}vel}}, \ and\ \bibinfo {author} {\bibfnamefont {E.}~\bibnamefont
  {Sch{\"o}ll}},\ }\bibfield  {title} {\enquote {\bibinfo {title} {Loss of
  coherence in dynamical networks: Spatial chaos and chimera states},}\
  }\href@noop {} {\bibfield  {journal} {\bibinfo  {journal} {Phys. Rev. Lett.}\
  }\textbf {\bibinfo {volume} {106}},\ \bibinfo {pages} {234102} (\bibinfo
  {year} {2011})}\BibitemShut {NoStop}%
\bibitem [{\citenamefont {Omel'chenko}\ \emph {et~al.}(2012)\citenamefont
  {Omel'chenko}, \citenamefont {Wolfrum}, \citenamefont {Yanchuk},
  \citenamefont {Maistrenko},\ and\ \citenamefont {Sudakov}}]{OWYMS:2012}%
  \BibitemOpen
  \bibfield  {author} {\bibinfo {author} {\bibfnamefont {O.~E.}\ \bibnamefont
  {Omel'chenko}}, \bibinfo {author} {\bibfnamefont {M.}~\bibnamefont
  {Wolfrum}}, \bibinfo {author} {\bibfnamefont {S.}~\bibnamefont {Yanchuk}},
  \bibinfo {author} {\bibfnamefont {Y.~L.}\ \bibnamefont {Maistrenko}}, \ and\
  \bibinfo {author} {\bibfnamefont {O.}~\bibnamefont {Sudakov}},\ }\bibfield
  {title} {\enquote {\bibinfo {title} {Stationary patterns of coherence and
  incoherence in two-dimensional arrays of non-locally-coupled phase
  oscillators},}\ }\href@noop {} {\bibfield  {journal} {\bibinfo  {journal}
  {Phys. Rev. E}\ }\textbf {\bibinfo {volume} {85}},\ \bibinfo {pages} {036210}
  (\bibinfo {year} {2012})}\BibitemShut {NoStop}%
\bibitem [{\citenamefont {Zhu}\ \emph {et~al.}(2012)\citenamefont {Zhu},
  \citenamefont {Li}, \citenamefont {Zhang},\ and\ \citenamefont
  {Yang}}]{ZLZY:2012}%
  \BibitemOpen
  \bibfield  {author} {\bibinfo {author} {\bibfnamefont {Y.}~\bibnamefont
  {Zhu}}, \bibinfo {author} {\bibfnamefont {Y.}~\bibnamefont {Li}}, \bibinfo
  {author} {\bibfnamefont {M.}~\bibnamefont {Zhang}}, \ and\ \bibinfo {author}
  {\bibfnamefont {J.}~\bibnamefont {Yang}},\ }\bibfield  {title} {\enquote
  {\bibinfo {title} {The oscillating two-cluster chimera state in non-locally
  coupled phase oscillators},}\ }\href@noop {} {\bibfield  {journal} {\bibinfo
  {journal} {EPL (Europhys. Lett.)}\ }\textbf {\bibinfo {volume} {97}},\
  \bibinfo {pages} {10009} (\bibinfo {year} {2012})}\BibitemShut {NoStop}%
\bibitem [{\citenamefont {Laing}\ \emph {et~al.}(2012)\citenamefont {Laing},
  \citenamefont {Rajendran},\ and\ \citenamefont {Kevrekidis}}]{LRK:2012}%
  \BibitemOpen
  \bibfield  {author} {\bibinfo {author} {\bibfnamefont {C.~R.}\ \bibnamefont
  {Laing}}, \bibinfo {author} {\bibfnamefont {K.}~\bibnamefont {Rajendran}}, \
  and\ \bibinfo {author} {\bibfnamefont {I.~G.}\ \bibnamefont {Kevrekidis}},\
  }\bibfield  {title} {\enquote {\bibinfo {title} {Chimeras in random
  non-complete networks of phase oscillators},}\ }\href@noop {} {\bibfield
  {journal} {\bibinfo  {journal} {Chaos}\ }\textbf {\bibinfo {volume} {22}},\
  \bibinfo {pages} {013132} (\bibinfo {year} {2012})}\BibitemShut {NoStop}%
\bibitem [{\citenamefont {Tinsley}\ \emph {et~al.}(2012)\citenamefont
  {Tinsley}, \citenamefont {Nkomo},\ and\ \citenamefont
  {Showalter}}]{TNS:2012}%
  \BibitemOpen
  \bibfield  {author} {\bibinfo {author} {\bibfnamefont {M.~R.}\ \bibnamefont
  {Tinsley}}, \bibinfo {author} {\bibfnamefont {S.}~\bibnamefont {Nkomo}}, \
  and\ \bibinfo {author} {\bibfnamefont {K.}~\bibnamefont {Showalter}},\
  }\bibfield  {title} {\enquote {\bibinfo {title} {Chimera and phase-cluster
  states in populations of coupled chemical oscillators},}\ }\href@noop {}
  {\bibfield  {journal} {\bibinfo  {journal} {Nat. Phys.}\ }\textbf {\bibinfo
  {volume} {8}},\ \bibinfo {pages} {662} (\bibinfo {year} {2012})}\BibitemShut
  {NoStop}%
\bibitem [{\citenamefont {Hagerstrom}\ \emph {et~al.}(2012)\citenamefont
  {Hagerstrom}, \citenamefont {Murphy}, \citenamefont {Roy}, \citenamefont
  {H\"{o}vel}, \citenamefont {Omelchenko},\ and\ \citenamefont
  {Sch{\"o}ll}}]{HMRHOS:2012}%
  \BibitemOpen
  \bibfield  {author} {\bibinfo {author} {\bibfnamefont {A.~M.}\ \bibnamefont
  {Hagerstrom}}, \bibinfo {author} {\bibfnamefont {T.~E.}\ \bibnamefont
  {Murphy}}, \bibinfo {author} {\bibfnamefont {R.}~\bibnamefont {Roy}},
  \bibinfo {author} {\bibfnamefont {P.}~\bibnamefont {H\"{o}vel}}, \bibinfo
  {author} {\bibfnamefont {I.}~\bibnamefont {Omelchenko}}, \ and\ \bibinfo
  {author} {\bibfnamefont {E.}~\bibnamefont {Sch{\"o}ll}},\ }\bibfield  {title}
  {\enquote {\bibinfo {title} {Experimental observation of chimeras in
  coupled-map lattices},}\ }\href@noop {} {\bibfield  {journal} {\bibinfo
  {journal} {Nat. Phys.}\ }\textbf {\bibinfo {volume} {8}},\ \bibinfo {pages}
  {658} (\bibinfo {year} {2012})}\BibitemShut {NoStop}%
\bibitem [{\citenamefont {Omelchenko}\ \emph {et~al.}(2013)\citenamefont
  {Omelchenko}, \citenamefont {Omel��chenko}, \citenamefont {H{\"o}vel},\
  and\ \citenamefont {Sch{\"o}ll}}]{OOHS:2013}%
  \BibitemOpen
  \bibfield  {author} {\bibinfo {author} {\bibfnamefont {I.}~\bibnamefont
  {Omelchenko}}, \bibinfo {author} {\bibfnamefont {E.}~\bibnamefont
  {Omel��chenko}}, \bibinfo {author} {\bibfnamefont {P.}~\bibnamefont
  {H{\"o}vel}}, \ and\ \bibinfo {author} {\bibfnamefont {E.}~\bibnamefont
  {Sch{\"o}ll}},\ }\bibfield  {title} {\enquote {\bibinfo {title} {When
  nonlocal coupling between oscillators becomes stronger: Patched synchrony or
  multichimera states},}\ }\href@noop {} {\bibfield  {journal} {\bibinfo
  {journal} {Phys. Rev. Lett.}\ }\textbf {\bibinfo {volume} {110}},\ \bibinfo
  {pages} {224101} (\bibinfo {year} {2013})}\BibitemShut {NoStop}%
\bibitem [{\citenamefont {Ujjwal}\ and\ \citenamefont
  {Ramaswamy}(2013)}]{UR:2013}%
  \BibitemOpen
  \bibfield  {author} {\bibinfo {author} {\bibfnamefont {S.~R.}\ \bibnamefont
  {Ujjwal}}\ and\ \bibinfo {author} {\bibfnamefont {R.}~\bibnamefont
  {Ramaswamy}},\ }\bibfield  {title} {\enquote {\bibinfo {title} {Chimeras with
  multiple coherent regions},}\ }\href@noop {} {\bibfield  {journal} {\bibinfo
  {journal} {Phys. Rev. E}\ }\textbf {\bibinfo {volume} {88}},\ \bibinfo
  {pages} {032902} (\bibinfo {year} {2013})}\BibitemShut {NoStop}%
\bibitem [{\citenamefont {Zhu}\ \emph {et~al.}(2013)\citenamefont {Zhu},
  \citenamefont {Zheng},\ and\ \citenamefont {Yang}}]{ZZY:2013}%
  \BibitemOpen
  \bibfield  {author} {\bibinfo {author} {\bibfnamefont {Y.}~\bibnamefont
  {Zhu}}, \bibinfo {author} {\bibfnamefont {Z.}~\bibnamefont {Zheng}}, \ and\
  \bibinfo {author} {\bibfnamefont {J.}~\bibnamefont {Yang}},\ }\bibfield
  {title} {\enquote {\bibinfo {title} {Reversed two-cluster chimera state in
  non-locally coupled oscillators with heterogeneous phase lags},}\ }\href@noop
  {} {\bibfield  {journal} {\bibinfo  {journal} {EPL (Europhys. Lett.)}\
  }\textbf {\bibinfo {volume} {103}},\ \bibinfo {pages} {10007} (\bibinfo
  {year} {2013})}\BibitemShut {NoStop}%
\bibitem [{\citenamefont {Yao}\ \emph {et~al.}(2013)\citenamefont {Yao},
  \citenamefont {Huang}, \citenamefont {Lai},\ and\ \citenamefont
  {Zheng}}]{YHLZ:2013}%
  \BibitemOpen
  \bibfield  {author} {\bibinfo {author} {\bibfnamefont {N.}~\bibnamefont
  {Yao}}, \bibinfo {author} {\bibfnamefont {Z.-G.}\ \bibnamefont {Huang}},
  \bibinfo {author} {\bibfnamefont {Y.-C.}\ \bibnamefont {Lai}}, \ and\
  \bibinfo {author} {\bibfnamefont {Z.-G.}\ \bibnamefont {Zheng}},\ }\bibfield
  {title} {\enquote {\bibinfo {title} {Robustness of chimera states in complex
  dynamical systems},}\ }\href@noop {} {\bibfield  {journal} {\bibinfo
  {journal} {Sci. Rep.}\ }\textbf {\bibinfo {volume} {3}},\ \bibinfo {pages}
  {3522} (\bibinfo {year} {2013})}\BibitemShut {NoStop}%
\bibitem [{\citenamefont {Nkomo}\ \emph {et~al.}(2013)\citenamefont {Nkomo},
  \citenamefont {Tinsley},\ and\ \citenamefont {Showalter}}]{NTS:2013}%
  \BibitemOpen
  \bibfield  {author} {\bibinfo {author} {\bibfnamefont {S.}~\bibnamefont
  {Nkomo}}, \bibinfo {author} {\bibfnamefont {M.~R.}\ \bibnamefont {Tinsley}},
  \ and\ \bibinfo {author} {\bibfnamefont {K.}~\bibnamefont {Showalter}},\
  }\bibfield  {title} {\enquote {\bibinfo {title} {Chimera states in
  populations of nonlocally coupled chemical oscillators},}\ }\href@noop {}
  {\bibfield  {journal} {\bibinfo  {journal} {Phys. Rev. Lett.}\ }\textbf
  {\bibinfo {volume} {110}},\ \bibinfo {pages} {244102} (\bibinfo {year}
  {2013})}\BibitemShut {NoStop}%
\bibitem [{\citenamefont {Martens}\ \emph {et~al.}(2013)\citenamefont
  {Martens}, \citenamefont {Thutupalli}, \citenamefont {Fourri{\`e}re},\ and\
  \citenamefont {Hallatschek}}]{MTFH:2013}%
  \BibitemOpen
  \bibfield  {author} {\bibinfo {author} {\bibfnamefont {E.~A.}\ \bibnamefont
  {Martens}}, \bibinfo {author} {\bibfnamefont {S.}~\bibnamefont {Thutupalli}},
  \bibinfo {author} {\bibfnamefont {A.}~\bibnamefont {Fourri{\`e}re}}, \ and\
  \bibinfo {author} {\bibfnamefont {O.}~\bibnamefont {Hallatschek}},\
  }\bibfield  {title} {\enquote {\bibinfo {title} {Chimera states in mechanical
  oscillator networks},}\ }\href@noop {} {\bibfield  {journal} {\bibinfo
  {journal} {Proc. Nat. Acad. Sci. (USA)}\ }\textbf {\bibinfo {volume} {110}},\
  \bibinfo {pages} {10563} (\bibinfo {year} {2013})}\BibitemShut {NoStop}%
\bibitem [{\citenamefont {Larger}\ \emph {et~al.}(2013)\citenamefont {Larger},
  \citenamefont {Penkovsky},\ and\ \citenamefont {Maistrenko}}]{LBM:2013}%
  \BibitemOpen
  \bibfield  {author} {\bibinfo {author} {\bibfnamefont {L.}~\bibnamefont
  {Larger}}, \bibinfo {author} {\bibfnamefont {B.}~\bibnamefont {Penkovsky}}, \
  and\ \bibinfo {author} {\bibfnamefont {Y.}~\bibnamefont {Maistrenko}},\
  }\bibfield  {title} {\enquote {\bibinfo {title} {Virtual chimera states for
  delayed-feedback systems},}\ }\href@noop {} {\bibfield  {journal} {\bibinfo
  {journal} {Phys. Rev. Lett.}\ }\textbf {\bibinfo {volume} {111}},\ \bibinfo
  {pages} {054103} (\bibinfo {year} {2013})}\BibitemShut {NoStop}%
\bibitem [{\citenamefont {Panaggio}\ and\ \citenamefont
  {Abrams}(2013)}]{PA:2013}%
  \BibitemOpen
  \bibfield  {author} {\bibinfo {author} {\bibfnamefont {M.~J.}\ \bibnamefont
  {Panaggio}}\ and\ \bibinfo {author} {\bibfnamefont {D.~M.}\ \bibnamefont
  {Abrams}},\ }\bibfield  {title} {\enquote {\bibinfo {title} {Chimera states
  on a flat torus},}\ }\href@noop {} {\bibfield  {journal} {\bibinfo  {journal}
  {Phys. Rev. Lett.}\ }\textbf {\bibinfo {volume} {110}},\ \bibinfo {pages}
  {094102} (\bibinfo {year} {2013})}\BibitemShut {NoStop}%
\bibitem [{\citenamefont {Gu}\ \emph {et~al.}(2013)\citenamefont {Gu},
  \citenamefont {St-Yves},\ and\ \citenamefont {Davidsen}}]{GSD:2013}%
  \BibitemOpen
  \bibfield  {author} {\bibinfo {author} {\bibfnamefont {C.}~\bibnamefont
  {Gu}}, \bibinfo {author} {\bibfnamefont {G.}~\bibnamefont {St-Yves}}, \ and\
  \bibinfo {author} {\bibfnamefont {J.}~\bibnamefont {Davidsen}},\ }\bibfield
  {title} {\enquote {\bibinfo {title} {Spiral wave chimeras in complex
  oscillatory and chaotic systems},}\ }\href@noop {} {\bibfield  {journal}
  {\bibinfo  {journal} {Phys. Rev. Lett.}\ }\textbf {\bibinfo {volume} {111}},\
  \bibinfo {pages} {134101} (\bibinfo {year} {2013})}\BibitemShut {NoStop}%
\bibitem [{\citenamefont {Sieber}\ \emph {et~al.}(2014)\citenamefont {Sieber},
  \citenamefont {Omel'chenko},\ and\ \citenamefont {Wolfrum}}]{SOW:2014}%
  \BibitemOpen
  \bibfield  {author} {\bibinfo {author} {\bibfnamefont {J.}~\bibnamefont
  {Sieber}}, \bibinfo {author} {\bibfnamefont {O.~E.}\ \bibnamefont
  {Omel'chenko}}, \ and\ \bibinfo {author} {\bibfnamefont {M.}~\bibnamefont
  {Wolfrum}},\ }\bibfield  {title} {\enquote {\bibinfo {title} {Controlling
  unstable chaos: Stabilizing chimera states by feedback},}\ }\href@noop {}
  {\bibfield  {journal} {\bibinfo  {journal} {Phys. Rev. Lett.}\ }\textbf
  {\bibinfo {volume} {112}},\ \bibinfo {pages} {054102} (\bibinfo {year}
  {2014})}\BibitemShut {NoStop}%
\bibitem [{\citenamefont {Schmidt}\ \emph {et~al.}(2014)\citenamefont
  {Schmidt}, \citenamefont {Sch{\"o}nleber}, \citenamefont {Krischer},\ and\
  \citenamefont {Garc{\'\i}a-Morales}}]{SSKG:2014}%
  \BibitemOpen
  \bibfield  {author} {\bibinfo {author} {\bibfnamefont {L.}~\bibnamefont
  {Schmidt}}, \bibinfo {author} {\bibfnamefont {K.}~\bibnamefont
  {Sch{\"o}nleber}}, \bibinfo {author} {\bibfnamefont {K.}~\bibnamefont
  {Krischer}}, \ and\ \bibinfo {author} {\bibfnamefont {V.}~\bibnamefont
  {Garc{\'\i}a-Morales}},\ }\bibfield  {title} {\enquote {\bibinfo {title}
  {Coexistence of synchrony and incoherence in oscillatory media under
  nonlinear global coupling},}\ }\href@noop {} {\bibfield  {journal} {\bibinfo
  {journal} {Chaos}\ }\textbf {\bibinfo {volume} {24}},\ \bibinfo {pages}
  {013102} (\bibinfo {year} {2014})}\BibitemShut {NoStop}%
\bibitem [{\citenamefont {Zhu}\ \emph {et~al.}(2014)\citenamefont {Zhu},
  \citenamefont {Zheng},\ and\ \citenamefont {Yang}}]{ZZY:2014}%
  \BibitemOpen
  \bibfield  {author} {\bibinfo {author} {\bibfnamefont {Y.}~\bibnamefont
  {Zhu}}, \bibinfo {author} {\bibfnamefont {Z.}~\bibnamefont {Zheng}}, \ and\
  \bibinfo {author} {\bibfnamefont {J.}~\bibnamefont {Yang}},\ }\bibfield
  {title} {\enquote {\bibinfo {title} {Chimera states on complex networks},}\
  }\href@noop {} {\bibfield  {journal} {\bibinfo  {journal} {Phys. Rev. E}\
  }\textbf {\bibinfo {volume} {89}},\ \bibinfo {pages} {022914} (\bibinfo
  {year} {2014})}\BibitemShut {NoStop}%
\bibitem [{\citenamefont {Omel'chenko}(2013)}]{O:2013}%
  \BibitemOpen
  \bibfield  {author} {\bibinfo {author} {\bibfnamefont {O.~E.}\ \bibnamefont
  {Omel'chenko}},\ }\bibfield  {title} {\enquote {\bibinfo {title}
  {Coherence-incoherence patterns in a ring of non-locally coupled phase
  oscillators},}\ }\href@noop {} {\bibfield  {journal} {\bibinfo  {journal}
  {Nonlinearity}\ }\textbf {\bibinfo {volume} {26}},\ \bibinfo {pages} {2469}
  (\bibinfo {year} {2013})}\BibitemShut {NoStop}%
\bibitem [{\citenamefont {Omel'chenko}\ \emph {et~al.}(2008)\citenamefont
  {Omel'chenko}, \citenamefont {Maistrenko},\ and\ \citenamefont
  {Tass}}]{OMT:2008}%
  \BibitemOpen
  \bibfield  {author} {\bibinfo {author} {\bibfnamefont {O.~E.}\ \bibnamefont
  {Omel'chenko}}, \bibinfo {author} {\bibfnamefont {Y.~L.}\ \bibnamefont
  {Maistrenko}}, \ and\ \bibinfo {author} {\bibfnamefont {P.~A.}\ \bibnamefont
  {Tass}},\ }\bibfield  {title} {\enquote {\bibinfo {title} {Chimera states:
  The natural link between coherence and incoherence},}\ }\href@noop {}
  {\bibfield  {journal} {\bibinfo  {journal} {Phys. Rev. Lett.}\ }\textbf
  {\bibinfo {volume} {100}},\ \bibinfo {pages} {044105} (\bibinfo {year}
  {2008})}\BibitemShut {NoStop}%
\bibitem [{\citenamefont {Xie}\ \emph {et~al.}(2014)\citenamefont {Xie},
  \citenamefont {Knobloch},\ and\ \citenamefont {Kao}}]{XKK:2014}%
  \BibitemOpen
  \bibfield  {author} {\bibinfo {author} {\bibfnamefont {J.}~\bibnamefont
  {Xie}}, \bibinfo {author} {\bibfnamefont {E.}~\bibnamefont {Knobloch}}, \
  and\ \bibinfo {author} {\bibfnamefont {H.-C.}\ \bibnamefont {Kao}},\
  }\bibfield  {title} {\enquote {\bibinfo {title} {Multicluster and traveling
  chimera states in nonlocal phase-coupled oscillators},}\ }\href@noop {}
  {\bibfield  {journal} {\bibinfo  {journal} {Phys. Rev. E}\ }\textbf {\bibinfo
  {volume} {90}},\ \bibinfo {pages} {022919} (\bibinfo {year}
  {2014})}\BibitemShut {NoStop}%
\bibitem [{\citenamefont {Yao}\ \emph {et~al.}(2015)\citenamefont {Yao},
  \citenamefont {Huang}, \citenamefont {Grebogi},\ and\ \citenamefont
  {Lai}}]{YHGL:2015}%
  \BibitemOpen
  \bibfield  {author} {\bibinfo {author} {\bibfnamefont {N.}~\bibnamefont
  {Yao}}, \bibinfo {author} {\bibfnamefont {Z.-G.}\ \bibnamefont {Huang}},
  \bibinfo {author} {\bibfnamefont {C.}~\bibnamefont {Grebogi}}, \ and\
  \bibinfo {author} {\bibfnamefont {Y.-C.}\ \bibnamefont {Lai}},\ }\bibfield
  {title} {\enquote {\bibinfo {title} {Emergence of multicluster chimera
  states},}\ }\href@noop {} {\bibfield  {journal} {\bibinfo  {journal} {Sci.
  Rep.}\ }\textbf {\bibinfo {volume} {5}},\ \bibinfo {pages} {12988} (\bibinfo
  {year} {2015})}\BibitemShut {NoStop}%
\bibitem [{\citenamefont {Maistrenko}\ \emph {et~al.}(2015)\citenamefont
  {Maistrenko}, \citenamefont {O.Sudakov}, \citenamefont {Osiv},\ and\
  \citenamefont {Maistrenko}}]{MSOM:2015}%
  \BibitemOpen
  \bibfield  {author} {\bibinfo {author} {\bibfnamefont {Y.}~\bibnamefont
  {Maistrenko}}, \bibinfo {author} {\bibnamefont {O.Sudakov}}, \bibinfo
  {author} {\bibfnamefont {O.}~\bibnamefont {Osiv}}, \ and\ \bibinfo {author}
  {\bibfnamefont {V.}~\bibnamefont {Maistrenko}},\ }\bibfield  {title}
  {\enquote {\bibinfo {title} {Chimera states in three dimensions},}\
  }\href@noop {} {\bibfield  {journal} {\bibinfo  {journal} {New J. Phys.}\
  }\textbf {\bibinfo {volume} {17}},\ \bibinfo {pages} {073037} (\bibinfo
  {year} {2015})}\BibitemShut {NoStop}%
\bibitem [{\citenamefont {Panaggio}\ and\ \citenamefont
  {Abrams}(2015{\natexlab{a}})}]{PA:2015}%
  \BibitemOpen
  \bibfield  {author} {\bibinfo {author} {\bibfnamefont {M.~J.}\ \bibnamefont
  {Panaggio}}\ and\ \bibinfo {author} {\bibfnamefont {D.~M.}\ \bibnamefont
  {Abrams}},\ }\bibfield  {title} {\enquote {\bibinfo {title} {Chimera states
  on the surface of a sphere},}\ }\href@noop {} {\bibfield  {journal} {\bibinfo
   {journal} {Phys. Rev. E.}\ }\textbf {\bibinfo {volume} {91}},\ \bibinfo
  {pages} {022909} (\bibinfo {year} {2015}{\natexlab{a}})}\BibitemShut
  {NoStop}%
\bibitem [{\citenamefont {Panaggio}\ and\ \citenamefont
  {Abrams}(2015{\natexlab{b}})}]{PAb:2015}%
  \BibitemOpen
  \bibfield  {author} {\bibinfo {author} {\bibfnamefont {M.~J.}\ \bibnamefont
  {Panaggio}}\ and\ \bibinfo {author} {\bibfnamefont {D.~M.}\ \bibnamefont
  {Abrams}},\ }\bibfield  {title} {\enquote {\bibinfo {title} {Chimera states:
  {Coexistence} of coherence and incoherence in networks of coupled
  oscillators},}\ }\href@noop {} {\bibfield  {journal} {\bibinfo  {journal}
  {Nonlinearity}\ }\textbf {\bibinfo {volume} {28}},\ \bibinfo {pages} {R67}
  (\bibinfo {year} {2015}{\natexlab{b}})}\BibitemShut {NoStop}%
\bibitem [{\citenamefont {Martens}\ and\ \citenamefont {Bick}(2015)}]{MB:2015}%
  \BibitemOpen
  \bibfield  {author} {\bibinfo {author} {\bibfnamefont {E.~A.}\ \bibnamefont
  {Martens}}\ and\ \bibinfo {author} {\bibfnamefont {C.}~\bibnamefont {Bick}},\
  }\bibfield  {title} {\enquote {\bibinfo {title} {Controlling chimeras},}\
  }\href@noop {} {\bibfield  {journal} {\bibinfo  {journal} {New J. Phys.}\
  }\textbf {\bibinfo {volume} {17}},\ \bibinfo {pages} {033030} (\bibinfo
  {year} {2015})}\BibitemShut {NoStop}%
\bibitem [{\citenamefont {Omelchenko}\ \emph
  {et~al.}(2015{\natexlab{a}})\citenamefont {Omelchenko}, \citenamefont
  {Zakharova}, \citenamefont {H\"{o}vel}, \citenamefont {Siebert},\ and\
  \citenamefont {Sch\"{o}ll}}]{OZHSS:2015}%
  \BibitemOpen
  \bibfield  {author} {\bibinfo {author} {\bibfnamefont {I.}~\bibnamefont
  {Omelchenko}}, \bibinfo {author} {\bibfnamefont {A.}~\bibnamefont
  {Zakharova}}, \bibinfo {author} {\bibfnamefont {P.}~\bibnamefont
  {H\"{o}vel}}, \bibinfo {author} {\bibfnamefont {J.}~\bibnamefont {Siebert}},
  \ and\ \bibinfo {author} {\bibfnamefont {E.}~\bibnamefont {Sch\"{o}ll}},\
  }\bibfield  {title} {\enquote {\bibinfo {title} {Nonlinearity of local
  dynamics promotes multi-chimeras},}\ }\href@noop {} {\bibfield  {journal}
  {\bibinfo  {journal} {Chaos}\ }\textbf {\bibinfo {volume} {25}},\ \bibinfo
  {pages} {083104} (\bibinfo {year} {2015}{\natexlab{a}})}\BibitemShut
  {NoStop}%
\bibitem [{\citenamefont {B\"ohm}\ \emph {et~al.}(2015)\citenamefont {B\"ohm},
  \citenamefont {Zakharova}, \citenamefont {Sch\"oll},\ and\ \citenamefont
  {L\"udge}}]{BZSL:2015}%
  \BibitemOpen
  \bibfield  {author} {\bibinfo {author} {\bibfnamefont {F.}~\bibnamefont
  {B\"ohm}}, \bibinfo {author} {\bibfnamefont {A.}~\bibnamefont {Zakharova}},
  \bibinfo {author} {\bibfnamefont {E.}~\bibnamefont {Sch\"oll}}, \ and\
  \bibinfo {author} {\bibfnamefont {K.}~\bibnamefont {L\"udge}},\ }\bibfield
  {title} {\enquote {\bibinfo {title} {Amplitude-phase coupling drives chimera
  states in globally coupled laser networks},}\ }\href {\doibase
  10.1103/PhysRevE.91.040901} {\bibfield  {journal} {\bibinfo  {journal} {Phys.
  Rev. E}\ }\textbf {\bibinfo {volume} {91}},\ \bibinfo {pages} {040901}
  (\bibinfo {year} {2015})}\BibitemShut {NoStop}%
\bibitem [{\citenamefont {Nkomo}\ \emph {et~al.}(2016)\citenamefont {Nkomo},
  \citenamefont {Tinsley},\ and\ \citenamefont {Showalter}}]{NTS:2016}%
  \BibitemOpen
  \bibfield  {author} {\bibinfo {author} {\bibfnamefont {S.}~\bibnamefont
  {Nkomo}}, \bibinfo {author} {\bibfnamefont {M.~R.}\ \bibnamefont {Tinsley}},
  \ and\ \bibinfo {author} {\bibfnamefont {K.}~\bibnamefont {Showalter}},\
  }\bibfield  {title} {\enquote {\bibinfo {title} {Chimera and chimera-like
  states in populations of nonlocally coupled homogeneous and heterogeneous
  chemical oscillators},}\ }\href@noop {} {\bibfield  {journal} {\bibinfo
  {journal} {Chaos}\ }\textbf {\bibinfo {volume} {26}},\ \bibinfo {pages}
  {094826} (\bibinfo {year} {2016})}\BibitemShut {NoStop}%
\bibitem [{\citenamefont {Viennot}\ and\ \citenamefont
  {Aubourg}(2016)}]{VA:2016}%
  \BibitemOpen
  \bibfield  {author} {\bibinfo {author} {\bibfnamefont {D.}~\bibnamefont
  {Viennot}}\ and\ \bibinfo {author} {\bibfnamefont {L.}~\bibnamefont
  {Aubourg}},\ }\bibfield  {title} {\enquote {\bibinfo {title} {Quantum chimera
  states},}\ }\href@noop {} {\bibfield  {journal} {\bibinfo  {journal} {Phys.
  Lett. A}\ }\textbf {\bibinfo {volume} {380}},\ \bibinfo {pages} {678}
  (\bibinfo {year} {2016})}\BibitemShut {NoStop}%
\bibitem [{\citenamefont {Hart}\ \emph {et~al.}(2016)\citenamefont {Hart},
  \citenamefont {Bansal}, \citenamefont {Murphy},\ and\ \citenamefont
  {Roy}}]{HBMR:2016}%
  \BibitemOpen
  \bibfield  {author} {\bibinfo {author} {\bibfnamefont {J.~D.}\ \bibnamefont
  {Hart}}, \bibinfo {author} {\bibfnamefont {K.}~\bibnamefont {Bansal}},
  \bibinfo {author} {\bibfnamefont {T.~E.}\ \bibnamefont {Murphy}}, \ and\
  \bibinfo {author} {\bibfnamefont {R.}~\bibnamefont {Roy}},\ }\bibfield
  {title} {\enquote {\bibinfo {title} {Experimental observation of chimera and
  cluster states in a minimal globally coupled network},}\ }\href@noop {}
  {\bibfield  {journal} {\bibinfo  {journal} {Chaos}\ }\textbf {\bibinfo
  {volume} {26}},\ \bibinfo {pages} {094801} (\bibinfo {year}
  {2016})}\BibitemShut {NoStop}%
\bibitem [{\citenamefont {Gambuzza}\ and\ \citenamefont
  {Frasca}(2016)}]{GF:2016}%
  \BibitemOpen
  \bibfield  {author} {\bibinfo {author} {\bibfnamefont {L.~V.}\ \bibnamefont
  {Gambuzza}}\ and\ \bibinfo {author} {\bibfnamefont {M.}~\bibnamefont
  {Frasca}},\ }\bibfield  {title} {\enquote {\bibinfo {title} {Pinning control
  of chimera states},}\ }\href {\doibase 10.1103/PhysRevE.94.022306} {\bibfield
   {journal} {\bibinfo  {journal} {Phys. Rev. E}\ }\textbf {\bibinfo {volume}
  {94}},\ \bibinfo {pages} {022306} (\bibinfo {year} {2016})}\BibitemShut
  {NoStop}%
\bibitem [{\citenamefont {Semenov}\ \emph {et~al.}(2016)\citenamefont
  {Semenov}, \citenamefont {Zakharova}, \citenamefont {Maistrenko},\ and\
  \citenamefont {Sch\"{o}ll}}]{SZMS:2016}%
  \BibitemOpen
  \bibfield  {author} {\bibinfo {author} {\bibfnamefont {V.}~\bibnamefont
  {Semenov}}, \bibinfo {author} {\bibfnamefont {A.}~\bibnamefont {Zakharova}},
  \bibinfo {author} {\bibfnamefont {Y.}~\bibnamefont {Maistrenko}}, \ and\
  \bibinfo {author} {\bibfnamefont {E.}~\bibnamefont {Sch\"{o}ll}},\ }\bibfield
   {title} {\enquote {\bibinfo {title} {Deterministic and stochastic control of
  chimera states in delayed feedback oscillator},}\ }\href@noop {} {\bibfield
  {journal} {\bibinfo  {journal} {AIP Conf. Proc.}\ }\textbf {\bibinfo {volume}
  {1738}},\ \bibinfo {pages} {210013} (\bibinfo {year} {2016})}\BibitemShut
  {NoStop}%
\bibitem [{\citenamefont {Kemeth}\ \emph {et~al.}(2016)\citenamefont {Kemeth},
  \citenamefont {Haugland}, \citenamefont {Schmidt}, \citenamefont
  {Kevrekidis},\ and\ \citenamefont {Krischer}}]{KHSKK:2016}%
  \BibitemOpen
  \bibfield  {author} {\bibinfo {author} {\bibfnamefont {F.~P.}\ \bibnamefont
  {Kemeth}}, \bibinfo {author} {\bibfnamefont {S.~W.}\ \bibnamefont
  {Haugland}}, \bibinfo {author} {\bibfnamefont {L.}~\bibnamefont {Schmidt}},
  \bibinfo {author} {\bibfnamefont {I.~G.}\ \bibnamefont {Kevrekidis}}, \ and\
  \bibinfo {author} {\bibfnamefont {K.}~\bibnamefont {Krischer}},\ }\bibfield
  {title} {\enquote {\bibinfo {title} {A classification scheme for chimera
  states},}\ }\href@noop {} {\bibfield  {journal} {\bibinfo  {journal} {Chaos}\
  }\textbf {\bibinfo {volume} {26}},\ \bibinfo {pages} {094815} (\bibinfo
  {year} {2016})}\BibitemShut {NoStop}%
\bibitem [{\citenamefont {Ulonskaa}\ \emph {et~al.}(2016)\citenamefont
  {Ulonskaa}, \citenamefont {Omelchenko}, \citenamefont {Zakharova},\ and\
  \citenamefont {Sch\"{o}ll}}]{UOZS:2016}%
  \BibitemOpen
  \bibfield  {author} {\bibinfo {author} {\bibfnamefont {S.}~\bibnamefont
  {Ulonskaa}}, \bibinfo {author} {\bibfnamefont {I.}~\bibnamefont
  {Omelchenko}}, \bibinfo {author} {\bibfnamefont {A.}~\bibnamefont
  {Zakharova}}, \ and\ \bibinfo {author} {\bibfnamefont {E.}~\bibnamefont
  {Sch\"{o}ll}},\ }\bibfield  {title} {\enquote {\bibinfo {title} {Chimera
  states in networks of van der {Pol} oscillators with hierarchical
  connectivities},}\ }\href@noop {} {\bibfield  {journal} {\bibinfo  {journal}
  {Chaos}\ }\textbf {\bibinfo {volume} {26}},\ \bibinfo {pages} {094825}
  (\bibinfo {year} {2016})}\BibitemShut {NoStop}%
\bibitem [{\citenamefont {Andrzejak}\ \emph {et~al.}(2017)\citenamefont
  {Andrzejak}, \citenamefont {Ruzzene},\ and\ \citenamefont
  {Malvestio}}]{ARM:2017}%
  \BibitemOpen
  \bibfield  {author} {\bibinfo {author} {\bibfnamefont {R.~G.}\ \bibnamefont
  {Andrzejak}}, \bibinfo {author} {\bibfnamefont {G.}~\bibnamefont {Ruzzene}},
  \ and\ \bibinfo {author} {\bibfnamefont {I.}~\bibnamefont {Malvestio}},\
  }\bibfield  {title} {\enquote {\bibinfo {title} {Generalized synchronization
  between chimera states},}\ }\href@noop {} {\bibfield  {journal} {\bibinfo
  {journal} {Chaos}\ }\textbf {\bibinfo {volume} {27}},\ \bibinfo {pages}
  {053114} (\bibinfo {year} {2017})}\BibitemShut {NoStop}%
\bibitem [{\citenamefont {Bera}\ \emph {et~al.}(2017)\citenamefont {Bera},
  \citenamefont {Majhi}, \citenamefont {Ghosh},\ and\ \citenamefont
  {Perc}}]{BMGP:2017}%
  \BibitemOpen
  \bibfield  {author} {\bibinfo {author} {\bibfnamefont {B.~K.}\ \bibnamefont
  {Bera}}, \bibinfo {author} {\bibfnamefont {S.}~\bibnamefont {Majhi}},
  \bibinfo {author} {\bibfnamefont {D.}~\bibnamefont {Ghosh}}, \ and\ \bibinfo
  {author} {\bibfnamefont {M.}~\bibnamefont {Perc}},\ }\bibfield  {title}
  {\enquote {\bibinfo {title} {Chimera states: Effects of different coupling
  topologies},}\ }\href@noop {} {\bibfield  {journal} {\bibinfo  {journal} {EPL
  (Europhys. Lett.)}\ }\textbf {\bibinfo {volume} {118}},\ \bibinfo {pages}
  {10001} (\bibinfo {year} {2017})}\BibitemShut {NoStop}%
\bibitem [{\citenamefont {Rakshit}\ \emph {et~al.}(2017)\citenamefont
  {Rakshit}, \citenamefont {Bera}, \citenamefont {Perc},\ and\ \citenamefont
  {Ghosh}}]{RBPG:2017}%
  \BibitemOpen
  \bibfield  {author} {\bibinfo {author} {\bibfnamefont {S.}~\bibnamefont
  {Rakshit}}, \bibinfo {author} {\bibfnamefont {B.~K.}\ \bibnamefont {Bera}},
  \bibinfo {author} {\bibfnamefont {M.}~\bibnamefont {Perc}}, \ and\ \bibinfo
  {author} {\bibfnamefont {D.}~\bibnamefont {Ghosh}},\ }\bibfield  {title}
  {\enquote {\bibinfo {title} {Basin stability for chimera states},}\
  }\href@noop {} {\bibfield  {journal} {\bibinfo  {journal} {Sci. Rep.}\
  }\textbf {\bibinfo {volume} {7}},\ \bibinfo {pages} {2412} (\bibinfo {year}
  {2017})}\BibitemShut {NoStop}%
\bibitem [{\citenamefont {Semenova}\ \emph {et~al.}(2017)\citenamefont
  {Semenova}, \citenamefont {Strelkova}, \citenamefont {Anishchenko},\ and\
  \citenamefont {Zakharova}}]{SSAZ:2017}%
  \BibitemOpen
  \bibfield  {author} {\bibinfo {author} {\bibfnamefont {N.~I.}\ \bibnamefont
  {Semenova}}, \bibinfo {author} {\bibfnamefont {G.~I.}\ \bibnamefont
  {Strelkova}}, \bibinfo {author} {\bibfnamefont {V.~S.}\ \bibnamefont
  {Anishchenko}}, \ and\ \bibinfo {author} {\bibfnamefont {A.}~\bibnamefont
  {Zakharova}},\ }\bibfield  {title} {\enquote {\bibinfo {title} {Temporal
  intermittency and the lifetime of chimera states in ensembles of nonlocally
  coupled chaotic oscillators},}\ }\href@noop {} {\bibfield  {journal}
  {\bibinfo  {journal} {Chaos}\ }\textbf {\bibinfo {volume} {27}},\ \bibinfo
  {pages} {061102} (\bibinfo {year} {2017})}\BibitemShut {NoStop}%
\bibitem [{\citenamefont {Malchow}\ \emph {et~al.}(2018)\citenamefont
  {Malchow}, \citenamefont {Omelchenko}, \citenamefont {Sch\"oll},\ and\
  \citenamefont {H\"ovel}}]{MOSH:2018}%
  \BibitemOpen
  \bibfield  {author} {\bibinfo {author} {\bibfnamefont {A.-K.}\ \bibnamefont
  {Malchow}}, \bibinfo {author} {\bibfnamefont {I.}~\bibnamefont {Omelchenko}},
  \bibinfo {author} {\bibfnamefont {E.}~\bibnamefont {Sch\"oll}}, \ and\
  \bibinfo {author} {\bibfnamefont {P.}~\bibnamefont {H\"ovel}},\ }\bibfield
  {title} {\enquote {\bibinfo {title} {Robustness of chimera states in
  nonlocally coupled networks of nonidentical logistic maps},}\ }\href
  {\doibase 10.1103/PhysRevE.98.012217} {\bibfield  {journal} {\bibinfo
  {journal} {Phys. Rev. E}\ }\textbf {\bibinfo {volume} {98}},\ \bibinfo
  {pages} {012217} (\bibinfo {year} {2018})}\BibitemShut {NoStop}%
\bibitem [{\citenamefont {Botha}\ and\ \citenamefont
  {Kolahchi}(2018)}]{BK:2018}%
  \BibitemOpen
  \bibfield  {author} {\bibinfo {author} {\bibfnamefont {A.~E.}\ \bibnamefont
  {Botha}}\ and\ \bibinfo {author} {\bibfnamefont {M.~R.}\ \bibnamefont
  {Kolahchi}},\ }\bibfield  {title} {\enquote {\bibinfo {title} {Analysis of
  chimera states as drive-response systems},}\ }\href@noop {} {\bibfield
  {journal} {\bibinfo  {journal} {Sci. Rep.}\ }\textbf {\bibinfo {volume}
  {8}},\ \bibinfo {pages} {1830} (\bibinfo {year} {2018})}\BibitemShut
  {NoStop}%
\bibitem [{\citenamefont {Omelchenko}\ \emph {et~al.}(2018)\citenamefont
  {Omelchenko}, \citenamefont {Omel'chenko}, \citenamefont {Zakharova},\ and\
  \citenamefont {Sch\"oll}}]{OOZS:2018}%
  \BibitemOpen
  \bibfield  {author} {\bibinfo {author} {\bibfnamefont {I.}~\bibnamefont
  {Omelchenko}}, \bibinfo {author} {\bibfnamefont {O.~E.}\ \bibnamefont
  {Omel'chenko}}, \bibinfo {author} {\bibfnamefont {A.}~\bibnamefont
  {Zakharova}}, \ and\ \bibinfo {author} {\bibfnamefont {E.}~\bibnamefont
  {Sch\"oll}},\ }\bibfield  {title} {\enquote {\bibinfo {title} {Optimal design
  of tweezer control for chimera states},}\ }\href {\doibase
  10.1103/PhysRevE.97.012216} {\bibfield  {journal} {\bibinfo  {journal} {Phys.
  Rev. E}\ }\textbf {\bibinfo {volume} {97}},\ \bibinfo {pages} {012216}
  (\bibinfo {year} {2018})}\BibitemShut {NoStop}%
\bibitem [{\citenamefont {Xu}\ \emph {et~al.}(2018)\citenamefont {Xu},
  \citenamefont {Wang}, \citenamefont {Huang},\ and\ \citenamefont
  {Lai}}]{XWHL:2018}%
  \BibitemOpen
  \bibfield  {author} {\bibinfo {author} {\bibfnamefont {H.-Y.}\ \bibnamefont
  {Xu}}, \bibinfo {author} {\bibfnamefont {G.-L.}\ \bibnamefont {Wang}},
  \bibinfo {author} {\bibfnamefont {L.}~\bibnamefont {Huang}}, \ and\ \bibinfo
  {author} {\bibfnamefont {Y.-C.}\ \bibnamefont {Lai}},\ }\bibfield  {title}
  {\enquote {\bibinfo {title} {Chaos in {Dirac} electron optics: Emergence of a
  relativistic quantum chimera},}\ }\href@noop {} {\bibfield  {journal}
  {\bibinfo  {journal} {Phys. Rev. Lett.}\ }\textbf {\bibinfo {volume} {120}},\
  \bibinfo {pages} {124101} (\bibinfo {year} {2018})}\BibitemShut {NoStop}%
\bibitem [{\citenamefont {Yao}\ \emph {et~al.}(2019)\citenamefont {Yao},
  \citenamefont {Huang}, \citenamefont {Ren}, \citenamefont {Grebogi},\ and\
  \citenamefont {Lai}}]{YHRGL:2019}%
  \BibitemOpen
  \bibfield  {author} {\bibinfo {author} {\bibfnamefont {N.}~\bibnamefont
  {Yao}}, \bibinfo {author} {\bibfnamefont {Z.-G.}\ \bibnamefont {Huang}},
  \bibinfo {author} {\bibfnamefont {H.-P.}\ \bibnamefont {Ren}}, \bibinfo
  {author} {\bibfnamefont {C.}~\bibnamefont {Grebogi}}, \ and\ \bibinfo
  {author} {\bibfnamefont {Y.-C.}\ \bibnamefont {Lai}},\ }\bibfield  {title}
  {\enquote {\bibinfo {title} {Self-adaptation of chimera states},}\ }\href
  {\doibase 10.1103/PhysRevE.99.010201} {\bibfield  {journal} {\bibinfo
  {journal} {Phys. Rev. E}\ }\textbf {\bibinfo {volume} {99}},\ \bibinfo
  {pages} {010201} (\bibinfo {year} {2019})}\BibitemShut {NoStop}%
\bibitem [{\citenamefont {Omelchenko}\ \emph
  {et~al.}(2015{\natexlab{b}})\citenamefont {Omelchenko}, \citenamefont
  {Provata}, \citenamefont {Hizanidis}, \citenamefont {Sch\"oll},\ and\
  \citenamefont {H\"ovel}}]{OPHSH:2015}%
  \BibitemOpen
  \bibfield  {author} {\bibinfo {author} {\bibfnamefont {I.}~\bibnamefont
  {Omelchenko}}, \bibinfo {author} {\bibfnamefont {A.}~\bibnamefont {Provata}},
  \bibinfo {author} {\bibfnamefont {J.}~\bibnamefont {Hizanidis}}, \bibinfo
  {author} {\bibfnamefont {E.}~\bibnamefont {Sch\"oll}}, \ and\ \bibinfo
  {author} {\bibfnamefont {P.}~\bibnamefont {H\"ovel}},\ }\bibfield  {title}
  {\enquote {\bibinfo {title} {Robustness of chimera states for coupled
  {FitzHugh-Nagumo} oscillators},}\ }\href {\doibase
  10.1103/PhysRevE.91.022917} {\bibfield  {journal} {\bibinfo  {journal} {Phys.
  Rev. E}\ }\textbf {\bibinfo {volume} {91}},\ \bibinfo {pages} {022917}
  (\bibinfo {year} {2015}{\natexlab{b}})}\BibitemShut {NoStop}%
\bibitem [{\citenamefont {Loos}\ \emph {et~al.}(2016)\citenamefont {Loos},
  \citenamefont {Claussen}, \citenamefont {Sch\"oll},\ and\ \citenamefont
  {Zakharova}}]{LCSZ:2016}%
  \BibitemOpen
  \bibfield  {author} {\bibinfo {author} {\bibfnamefont {S.~A.~M.}\
  \bibnamefont {Loos}}, \bibinfo {author} {\bibfnamefont {J.~C.}\ \bibnamefont
  {Claussen}}, \bibinfo {author} {\bibfnamefont {E.}~\bibnamefont {Sch\"oll}},
  \ and\ \bibinfo {author} {\bibfnamefont {A.}~\bibnamefont {Zakharova}},\
  }\bibfield  {title} {\enquote {\bibinfo {title} {Chimera patterns under the
  impact of noise},}\ }\href {\doibase 10.1103/PhysRevE.93.012209} {\bibfield
  {journal} {\bibinfo  {journal} {Phys. Rev. E}\ }\textbf {\bibinfo {volume}
  {93}},\ \bibinfo {pages} {012209} (\bibinfo {year} {2016})}\BibitemShut
  {NoStop}%
\bibitem [{\citenamefont {Olfati-Saber}(2006)}]{Olfati:2006}%
  \BibitemOpen
  \bibfield  {author} {\bibinfo {author} {\bibfnamefont {R.}~\bibnamefont
  {Olfati-Saber}},\ }\bibfield  {title} {\enquote {\bibinfo {title} {Swarms on
  sphere: A programmable swarm with synchronous behaviors like oscillator
  networks},}\ }in\ \href@noop {} {\emph {\bibinfo {booktitle} {Proceedings of
  the 45th IEEE Conference on Decision and Control}}}\ (\bibinfo {organization}
  {IEEE},\ \bibinfo {year} {2006})\ pp.\ \bibinfo {pages}
  {5060--5066}\BibitemShut {NoStop}%
\bibitem [{\citenamefont {Lohe}(2009)}]{Lohe:2009}%
  \BibitemOpen
  \bibfield  {author} {\bibinfo {author} {\bibfnamefont {M.}~\bibnamefont
  {Lohe}},\ }\bibfield  {title} {\enquote {\bibinfo {title} {{Non-Abelian
  Kuramoto} models and synchronization},}\ }\href@noop {} {\bibfield  {journal}
  {\bibinfo  {journal} {J. Phys. A Math. Theo.}\ }\textbf {\bibinfo {volume}
  {42}},\ \bibinfo {pages} {395101} (\bibinfo {year} {2009})}\BibitemShut
  {NoStop}%
\bibitem [{\citenamefont {Chandra}\ \emph
  {et~al.}(2019{\natexlab{a}})\citenamefont {Chandra}, \citenamefont {Girvan},\
  and\ \citenamefont {Ott}}]{CGO:2019}%
  \BibitemOpen
  \bibfield  {author} {\bibinfo {author} {\bibfnamefont {S.}~\bibnamefont
  {Chandra}}, \bibinfo {author} {\bibfnamefont {M.}~\bibnamefont {Girvan}}, \
  and\ \bibinfo {author} {\bibfnamefont {E.}~\bibnamefont {Ott}},\ }\bibfield
  {title} {\enquote {\bibinfo {title} {Continuous versus discontinuous
  transitions in the {D-dimensional generalized Kuramoto model: Odd D is
  different}},}\ }\href@noop {} {\bibfield  {journal} {\bibinfo  {journal}
  {Phys. Rev. X}\ }\textbf {\bibinfo {volume} {9}},\ \bibinfo {pages} {011002}
  (\bibinfo {year} {2019}{\natexlab{a}})}\BibitemShut {NoStop}%
\bibitem [{\citenamefont {Chandra}\ \emph
  {et~al.}(2019{\natexlab{b}})\citenamefont {Chandra}, \citenamefont {Girvan},\
  and\ \citenamefont {Ott}}]{CGO:2019b}%
  \BibitemOpen
  \bibfield  {author} {\bibinfo {author} {\bibfnamefont {S.}~\bibnamefont
  {Chandra}}, \bibinfo {author} {\bibfnamefont {M.}~\bibnamefont {Girvan}}, \
  and\ \bibinfo {author} {\bibfnamefont {E.}~\bibnamefont {Ott}},\ }\bibfield
  {title} {\enquote {\bibinfo {title} {Complexity reduction ansatz for systems
  of interacting orientable agents: Beyond the {Kuramoto} model},}\ }\href@noop
  {} {\bibfield  {journal} {\bibinfo  {journal} {Chaos}\ }\textbf {\bibinfo
  {volume} {29}},\ \bibinfo {pages} {053107} (\bibinfo {year}
  {2019}{\natexlab{b}})}\BibitemShut {NoStop}%
\bibitem [{\citenamefont {Grebogi}\ \emph {et~al.}(1983)\citenamefont
  {Grebogi}, \citenamefont {Ott},\ and\ \citenamefont {Yorke}}]{GOY:1983}%
  \BibitemOpen
  \bibfield  {author} {\bibinfo {author} {\bibfnamefont {C.}~\bibnamefont
  {Grebogi}}, \bibinfo {author} {\bibfnamefont {E.}~\bibnamefont {Ott}}, \ and\
  \bibinfo {author} {\bibfnamefont {J.~A.}\ \bibnamefont {Yorke}},\ }\bibfield
  {title} {\enquote {\bibinfo {title} {Crises, sudden changes in chaotic
  attractors and chaotic transients},}\ }\href@noop {} {\bibfield  {journal}
  {\bibinfo  {journal} {Physica D}\ }\textbf {\bibinfo {volume} {7}},\ \bibinfo
  {pages} {181} (\bibinfo {year} {1983})}\BibitemShut {NoStop}%
\bibitem [{\citenamefont {Lai}\ and\ \citenamefont {T{\'e}l}(2011)}]{LT:book}%
  \BibitemOpen
  \bibfield  {author} {\bibinfo {author} {\bibfnamefont {Y.-C.}\ \bibnamefont
  {Lai}}\ and\ \bibinfo {author} {\bibfnamefont {T.}~\bibnamefont {T{\'e}l}},\
  }\href@noop {} {\emph {\bibinfo {title} {Transient Chaos: Complex Dynamics on
  Finite Time Scales}}},\ Vol.\ \bibinfo {volume} {173}\ (\bibinfo  {publisher}
  {Springer},\ \bibinfo {address} {New York},\ \bibinfo {year}
  {2011})\BibitemShut {NoStop}%
\bibitem [{\citenamefont {Crutchfield}\ and\ \citenamefont
  {Kaneko}(1988)}]{CK:1988}%
  \BibitemOpen
  \bibfield  {author} {\bibinfo {author} {\bibfnamefont {J.~R.}\ \bibnamefont
  {Crutchfield}}\ and\ \bibinfo {author} {\bibfnamefont {K.}~\bibnamefont
  {Kaneko}},\ }\bibfield  {title} {\enquote {\bibinfo {title} {Are attractors
  relevant to turbulence?}}\ }\href@noop {} {\bibfield  {journal} {\bibinfo
  {journal} {Phys. Rev. Lett.}\ }\textbf {\bibinfo {volume} {60}},\ \bibinfo
  {pages} {2715} (\bibinfo {year} {1988})}\BibitemShut {NoStop}%
\bibitem [{\citenamefont {Grebogi}\ \emph {et~al.}(1985)\citenamefont
  {Grebogi}, \citenamefont {Ott},\ and\ \citenamefont {Yorke}}]{GOY:1985}%
  \BibitemOpen
  \bibfield  {author} {\bibinfo {author} {\bibfnamefont {C.}~\bibnamefont
  {Grebogi}}, \bibinfo {author} {\bibfnamefont {E.}~\bibnamefont {Ott}}, \ and\
  \bibinfo {author} {\bibfnamefont {J.}~\bibnamefont {Yorke}},\ }\bibfield
  {title} {\enquote {\bibinfo {title} {Super persistent chaotic transients},}\
  }\href@noop {} {\bibfield  {journal} {\bibinfo  {journal} {Ergod. Theor. Dyn.
  Syst.}\ }\textbf {\bibinfo {volume} {5}},\ \bibinfo {pages} {341} (\bibinfo
  {year} {1985})}\BibitemShut {NoStop}%
\bibitem [{\citenamefont {Do}\ and\ \citenamefont {Lai}(2003)}]{DL:2003}%
  \BibitemOpen
  \bibfield  {author} {\bibinfo {author} {\bibfnamefont {Y.}~\bibnamefont
  {Do}}\ and\ \bibinfo {author} {\bibfnamefont {Y.-C.}\ \bibnamefont {Lai}},\
  }\bibfield  {title} {\enquote {\bibinfo {title} {Superpersistent chaotic
  transients in physical space: advective dynamics of inertial particles in
  open chaotic flows under noise},}\ }\href@noop {} {\bibfield  {journal}
  {\bibinfo  {journal} {Phys. Rev. Lett.}\ }\textbf {\bibinfo {volume} {91}},\
  \bibinfo {pages} {224101} (\bibinfo {year} {2003})}\BibitemShut {NoStop}%
\bibitem [{\citenamefont {Do}\ and\ \citenamefont {Lai}(2004)}]{DL:2004}%
  \BibitemOpen
  \bibfield  {author} {\bibinfo {author} {\bibfnamefont {Y.}~\bibnamefont
  {Do}}\ and\ \bibinfo {author} {\bibfnamefont {Y.-C.}\ \bibnamefont {Lai}},\
  }\bibfield  {title} {\enquote {\bibinfo {title} {Extraordinarily
  superpersistent chaotic transients},}\ }\href@noop {} {\bibfield  {journal}
  {\bibinfo  {journal} {EPL (Europhys. Lett.)}\ }\textbf {\bibinfo {volume}
  {67}},\ \bibinfo {pages} {914} (\bibinfo {year} {2004})}\BibitemShut
  {NoStop}%
\bibitem [{\citenamefont {D'Huys}\ \emph {et~al.}(2016)\citenamefont {D'Huys},
  \citenamefont {Lohmann}, \citenamefont {Haynes},\ and\ \citenamefont
  {Gauthier}}]{DLHG:2016}%
  \BibitemOpen
  \bibfield  {author} {\bibinfo {author} {\bibfnamefont {O.}~\bibnamefont
  {D'Huys}}, \bibinfo {author} {\bibfnamefont {J.}~\bibnamefont {Lohmann}},
  \bibinfo {author} {\bibfnamefont {N.~D.}\ \bibnamefont {Haynes}}, \ and\
  \bibinfo {author} {\bibfnamefont {D.~J.}\ \bibnamefont {Gauthier}},\
  }\bibfield  {title} {\enquote {\bibinfo {title} {Super-transient scaling in
  time-delay autonomous boolean network motifs},}\ }\href@noop {} {\bibfield
  {journal} {\bibinfo  {journal} {Chaos}\ }\textbf {\bibinfo {volume} {26}},\
  \bibinfo {pages} {094810} (\bibinfo {year} {2016})}\BibitemShut {NoStop}%
\bibitem [{\citenamefont {Lohmann}\ \emph {et~al.}(2017)\citenamefont
  {Lohmann}, \citenamefont {D'Huys}, \citenamefont {Haynes}, \citenamefont
  {Sch\"oll},\ and\ \citenamefont {Gauthier}}]{LDHSG:2017}%
  \BibitemOpen
  \bibfield  {author} {\bibinfo {author} {\bibfnamefont {J.}~\bibnamefont
  {Lohmann}}, \bibinfo {author} {\bibfnamefont {O.}~\bibnamefont {D'Huys}},
  \bibinfo {author} {\bibfnamefont {N.~D.}\ \bibnamefont {Haynes}}, \bibinfo
  {author} {\bibfnamefont {E.}~\bibnamefont {Sch\"oll}}, \ and\ \bibinfo
  {author} {\bibfnamefont {D.~J.}\ \bibnamefont {Gauthier}},\ }\bibfield
  {title} {\enquote {\bibinfo {title} {Transient dynamics and their control in
  time-delay autonomous boolean ring networks},}\ }\href {\doibase
  10.1103/PhysRevE.95.022211} {\bibfield  {journal} {\bibinfo  {journal} {Phys.
  Rev. E}\ }\textbf {\bibinfo {volume} {95}},\ \bibinfo {pages} {022211}
  (\bibinfo {year} {2017})}\BibitemShut {NoStop}%
\bibitem [{\citenamefont {Rosin}\ \emph {et~al.}(2014)\citenamefont {Rosin},
  \citenamefont {Rontani}, \citenamefont {Haynes}, \citenamefont {Sch\"oll},\
  and\ \citenamefont {Gauthier}}]{RRHSG:2014}%
  \BibitemOpen
  \bibfield  {author} {\bibinfo {author} {\bibfnamefont {D.~P.}\ \bibnamefont
  {Rosin}}, \bibinfo {author} {\bibfnamefont {D.}~\bibnamefont {Rontani}},
  \bibinfo {author} {\bibfnamefont {N.~D.}\ \bibnamefont {Haynes}}, \bibinfo
  {author} {\bibfnamefont {E.}~\bibnamefont {Sch\"oll}}, \ and\ \bibinfo
  {author} {\bibfnamefont {D.~J.}\ \bibnamefont {Gauthier}},\ }\bibfield
  {title} {\enquote {\bibinfo {title} {Transient scaling and resurgence of
  chimera states in networks of boolean phase oscillators},}\ }\href {\doibase
  10.1103/PhysRevE.90.030902} {\bibfield  {journal} {\bibinfo  {journal} {Phys.
  Rev. E}\ }\textbf {\bibinfo {volume} {90}},\ \bibinfo {pages} {030902}
  (\bibinfo {year} {2014})}\BibitemShut {NoStop}%
\bibitem [{\citenamefont {Olmi}(2015)}]{Olmi:2015}%
  \BibitemOpen
  \bibfield  {author} {\bibinfo {author} {\bibfnamefont {S.}~\bibnamefont
  {Olmi}},\ }\bibfield  {title} {\enquote {\bibinfo {title} {Chimera states in
  coupled {Kuramoto} oscillators with inertia},}\ }\href@noop {} {\bibfield
  {journal} {\bibinfo  {journal} {Chaos}\ }\textbf {\bibinfo {volume} {25}},\
  \bibinfo {pages} {123125} (\bibinfo {year} {2015})}\BibitemShut {NoStop}%
\bibitem [{\citenamefont {Laing}(2019)}]{Laing:2019}%
  \BibitemOpen
  \bibfield  {author} {\bibinfo {author} {\bibfnamefont {C.~R.}\ \bibnamefont
  {Laing}},\ }\bibfield  {title} {\enquote {\bibinfo {title} {Dynamics and
  stability of chimera states in two coupled populations of oscillators},}\
  }\href {\doibase 10.1103/PhysRevE.100.042211} {\bibfield  {journal} {\bibinfo
   {journal} {Phys. Rev. E}\ }\textbf {\bibinfo {volume} {100}},\ \bibinfo
  {pages} {042211} (\bibinfo {year} {2019})}\BibitemShut {NoStop}%
\bibitem [{\citenamefont {Ott}\ and\ \citenamefont {Antonsen}(2008)}]{OA:2008}%
  \BibitemOpen
  \bibfield  {author} {\bibinfo {author} {\bibfnamefont {E.}~\bibnamefont
  {Ott}}\ and\ \bibinfo {author} {\bibfnamefont {T.~M.}\ \bibnamefont
  {Antonsen}},\ }\bibfield  {title} {\enquote {\bibinfo {title} {Low
  dimensional behavior of large systems of globally coupled oscillators},}\
  }\href@noop {} {\bibfield  {journal} {\bibinfo  {journal} {Chaos}\ }\textbf
  {\bibinfo {volume} {18}},\ \bibinfo {pages} {037113} (\bibinfo {year}
  {2008})}\BibitemShut {NoStop}%
\bibitem [{\citenamefont {Ott}\ and\ \citenamefont {Antonsen}(2009)}]{OA:2009}%
  \BibitemOpen
  \bibfield  {author} {\bibinfo {author} {\bibfnamefont {E.}~\bibnamefont
  {Ott}}\ and\ \bibinfo {author} {\bibfnamefont {T.~M.}\ \bibnamefont
  {Antonsen}},\ }\bibfield  {title} {\enquote {\bibinfo {title} {Long time
  evolution of phase oscillator systems},}\ }\href@noop {} {\bibfield
  {journal} {\bibinfo  {journal} {Chaos}\ }\textbf {\bibinfo {volume} {19}},\
  \bibinfo {pages} {023117} (\bibinfo {year} {2009})}\BibitemShut {NoStop}%
\end{thebibliography}

%
\end{document}